 \documentclass[final,5p,times,twocolumn,authoryear]{elsarticle}

\usepackage{amssymb}
\usepackage{lipsum}
\usepackage{amsmath}
\usepackage{xcolor}
\usepackage{booktabs}
\usepackage{multirow}
\usepackage{textcomp} 
\usepackage{float}
\usepackage[colorlinks=true,linkcolor=blue,citecolor=blue,urlcolor=blue]{hyperref}

\usepackage{marvosym}

\journal{High Energy Astrophysics}

\begin{document}

\begin{frontmatter}



\title{A set of distinctive properties ruling the prompt emission of GRB\,230307A and other long $\gamma$-ray bursts from compact object mergers}

\author[unife,inafbo]{R.~Maccary}\ead{romain.maccary@edu.unife.it}
\author[unife,inafbo,infnfe]{C.~Guidorzi}
\author[unife,inafbo]{M.~Maistrello}
\author[ljmu]{S.~Kobayashi}
\author[unife,infnfe,inafabruzzo]{M.~Bulla}
\author[ihep]{R.~Moradi}
\author[ihep]{S.-X.~Yi}
\author[ihep,cas]{C.~W.~Wang}
\author[ihep,qufu]{W.~L.~Zhang}
\author[ihep,cas]{W.-J.~Tan}
\author[ihep]{S.-L~Xiong}
\author[ihep,cas]{S.-N.~Zhang}


\affiliation[unife]{organization={Department of Physics and Earth Science, University of Ferrara}, addressline={via Saragat 1}, postcode={I--44122}, city={Ferrara}, country={Italy}}

\affiliation[inafbo]{organization={INAF -- Osservatorio di Astrofisica e Scienza dello Spazio di Bologna}, addressline={Via Piero Gobetti 101}, postcode={I-40129}, city={Bologna}, country={Italy}}

\affiliation[infnfe]{organization={INFN -- Sezione di Ferrara}, addressline={via Saragat 1}, postcode={I--44122}, city={Ferrara}, country={Italy}}

\affiliation[inafabruzzo]{organization={INAF, Osservatorio Astronomico d’Abruzzo}, addressline={Via Mentore Maggini snc},postcode={64100 Teramo},country={Italy}}

\affiliation[ljmu]{organization={Astrophysics Research Institute, Liverpool John Moores University}, addressline={Liverpool Science Park IC2, 146 Brownlow Hill}, city={Liverpool},  postcode={L3 5RF}, country={UK}}

\affiliation[ihep]{organization={State Key Laboratory for Particle Astrophysics, Institute of High Energy Physics, Chinese Academy of Sciences},addressline={19B Yuquan Road},postcode= {Beijing 100049}, country={China}}

\affiliation[cas]{organization={University of Chinese Academy of Sciences, Chinese Academy of Sciences}, addressline={ Beijing 100049}, country={China}}

\affiliation[qufu]{organization={School of Physics and Physical Engineering, Qufu Normal University}, addresline={Qufu, Shandong 273165}, country={China}}
\begin{abstract}
Short gamma-ray bursts (SGRBs), occasionally followed by a long and spectrally soft extended emission, are associated with compact object mergers (COMs). Yet, a few recent long GRBs (LGRBs) show compelling evidence for a COM origin, in contrast with the massive-star core-collapse origin of most LGRBs. While possible COM indicators were found, such as the minimum variability timescale (MVT), a detailed and unique characterisation of their $\gamma$-ray prompt emission that may help identify and explain their deceptively long profile is yet to be found.
Here we report the discovery of a set of distinctive properties that rule the temporal and spectral evolution of GRB\,230307A, a LGRB with evidence for a COM origin. Specifically, the sequence of pulses that make up its profile is characterised by an exponential evolution of (i) flux intensities, (ii) waiting times between adjacent pulses, (iii) pulse durations, and (iv) spectral peak energy. Analogous patterns are observed in the prompt emission of other long COM candidates. The observed evolution of gamma-ray pulses {would imply} that a relativistic jet is colliding with more slowly expanding material.
This contrasts with the standard internal shock model for typical LGRBs, in which dissipation occurs at random locations within the jet itself.
We tentatively propose a few simple toy models that may explain these properties and are able to reproduce the overall time profile.

\end{abstract}




\end{frontmatter}




\section{Introduction}
\label{sec:intro}
At least two kinds of progenitors of gamma-ray bursts (GRBs) are known: (i) the core-collapse of some kind of  massive stars (a collapsar; \citealt{Woosley93,Paczynski98,MacFadyen99}); (ii) compact object mergers (COMs), in particular the coalescence of two neutron stars (NS; \citealt{Eichler89,Paczynski91,Narayan92}),  or a NS with a black hole (BH). While the former class usually leads to a long ($\gtrsim 2$~s) GRB, the latter typically results in a short and hard GRB and, in the optical band, is expected to be associated with a kilonova (KN), radiation powered by the radioactive decay of $r$-process elements created in the aftermath of the merger (see \citealt{Metzger20_revKN} for a review).

Although GRB duration was initially considered as an irrefutable property revealing the progenitor's nature, a number of baffling cases have recently been discovered: the apparently short GRB\,200826A \citep{Ahumada21,Zhang21a,Rossi22a}, which is instead associated with a core-collapse supernova (SN), or long GRBs (LGRBs) GRB\,060614 \citep{DellaValle06,Fynbo06,Jin15}, GRB\,211211A \citep{Rastinejad22,Yang22,Troja22,Xiao22}, GRB\,191019A \citep{Levan23,Stratta25}, for which a COM origin is strongly favoured. Furthermore, a subclass of events emerged —known as short GRBs with extended emission (SEE-GRBs)— which are characterised by an initial narrow, hard spike followed by a longer, softer tail \citep{Norris06}. These events, possibly originating from mergers, further challenged the reliability of GRB duration as a robust classification criterion. To avoid confusion, \citet{Zhang09c} referred to GRBs with a merger origin as Type I GRBs, and to collapsars as Type II GRBs.

Recent GRB\,230307A was thrust into the spotlight thanks to its exceptional brightness along with its peculiar properties. On March 7, 2023, at 15:44:06.650 UTC, GRB\,230307A
triggered the Gravitational wave high-energy Electromagnetic
Counterpart All-sky Monitor (GECAM; \citealt{Li20}), which
observed it without saturation effects \citep{Xiong23}. Despite its long duration ($T_{\rm 90} \sim  30~{\rm s}$), this burst shows several pieces of evidence for a COM origin. Indeed, its locations in the Amati \citep{Amati02,Amati06} and Yonetoku planes \citep{Yonetoku04} fall outside the $90\%$ confidence predictions for Type II GRBs \citep{Svinkin23_230307A}, and a short minimum variability timescale (MVT) of 
around 30 ms—typical of Type I GRBs—is observed \citep{Camisasca23b}. Moreover, the X-ray afterglow flux, rescaled by the early $\gamma$-ray emission fluence, is comparably faint to those of other long Type I GRBs. Most importantly, the optical transient associated with it and with a projected offset of $30~{\rm  kpc}$ from its host galaxy, showed photometric and spectroscopic evidence for the presence of a KN \citep{Levan24,Gillanders25}, which tipped the balance towards a COM origin.
   
GRB\,230307A exhibits a well-structured $\gamma$ -ray light curve (LC), made of three distinct episodes: a short initial soft spike, identified as a 
precursor \citep{Dichiara23}, shortly followed by a long and hard main emission preceding a softer extended emission, the main and the extended emission being separated by a dip-like feature. This three-phase structure has been claimed to be shared by some similar long Type I GRBs, possibly forming a sub-class of the Type I GRBs, referred to as Type IL GRBs (e.g \citealt{Wang25,Tan25}).
GRB\,230307A shares many properties with GRB\,211211A: both exploded at close distances ($z=0.076$ for GRB\,211211A and $z=0.065$ for GRB\,230307A), have short MVTs of a few ten ms, and are clear outliers of the Amati relation for Type II GRBs \citep{Peng24}.

NS-NS and NS-BH mergers might struggle to produce GRBs as long as GRB\,230307A, owing to their short accretion timescale. A white dwarf (WD) disrupted by a NS companion, 
producing a less compact remnant, might be able to last long enough to explain the GRB duration. \citet{Wang24} argues that a WD-NS merger could lead to a magnetar that could power a long GRB and a KN. The soft X-ray light curve measured by the Lobster Eye Imager for Astronomy (LEIA; \citealt{Zhang22}) presents a plateau followed by a steeper decay compatible with a magnetar spin-down model. \citet{Sun25} shows that broad-band (soft X to $\gamma$-rays) observations of GRB\,230307A revealed a distinct X-ray component, possibly due to a newly born magnetar. A magnetar as a GRB central engine would lead to a Poynting-flux dominated outflow, which is confirmed by the non-detection of a thermal component, implying a high magnetisation parameter ($\sigma > 7$ at radius $R_0=10^{10}~{\rm cm}$) to suppress the expected photospheric emission \citep{Du24b}. 

The emission mechanism powering GRB\,230307A is uncertain. \citet{Moradi24} carried out a systematic spectral and
temporal analysis of \textit{Fermi}/GBM and GECAM data and showed that the energy flux and the peak energy temporal evolution in the late prompt emission are not fully compatible with the predictions of the internal shock (IS) model \citep{Rees94,Kobayashi97,Daigne98}, although uncertainties in the micro-physics of the shock region might be responsible for the observed discrepancies \citep{Bosnjak14b}. \citet{Yi25b} shows that the long broad pulse shaping the overall GRB\,230307A time profile could be the result of many superimposed narrow pulses produced by local magnetic reconnection events, as foreseen by the Internal-Collision-Induced Magnetic Reconnection and Turbulence (ICMART; \citealt{ICMART}) model. \citet{Yi25} proposed a model in which a brief energy injection from the central engine triggers turbulence in a small localised region. Turbulence then propagates radially as the jet expands forward, moving away from the central engine. This model predicts a single broad pulse that widens with decreasing energy ranges and progressively softens throughout the burst.

A progressive increase of the pulse width over time is a natural outcome of external shocks (ES). An intense debate emerged among the GRB community  in the late 90s as to whether the dissipation mechanism into $\gamma$-rays was due to internal or external shocks \citep{Fenimore96,Dermer99,Dermer08}. The absence of any timescale evolution along the profile of GRB\,990123 \citep{Fenimore99} and of other long and multi-peaked GRBs tipped the balance towards ISs. In this respect, the temporal evolution of GRB\,230307A, as well as of other similar events, might suggest a new distinctive hallmark of the elusive class of LGRBs that are COM candidates.

In this work, we focus on the GECAM LC in the 30-6000 keV passband: owing to the exquisite quality of the data, we could identify about one hundred peaks, thus enabling a  statistical analysis, which is usually impossible for a single GRB. As a result, the prompt emission of GRB\,230307A displays a set of distinctive properties, which suggest a different origin from the canonical LGRBs associated with the core-collapse of massive stars. Some of these properties are also observed in similar long-duration COM candidates mentioned above, as revealed by a preliminary analysis.

The identification and characterisation of this rare set of observed properties, which appear to distinguish COM candidates from most long GRBs, and particularly from collapsars, form the core of the present work. As a secondary objective, we tentatively propose a few possible toy models and interpretations that aim to account for these properties in a self-consistent way, discussing the advantages and limitations of each, without necessarily excluding alternative scenarios.
Data analysis is reported in Section \ref{sec:data_anal}, results are presented in Section \ref{sec:res} and discussed in Section \ref{sec:disc}. Conclusions are drawn in Section \ref{sec:conc}. We adopted the cosmological parameter values from \citet{cosmoPlanck20}.

\section{Data analysis}
\label{sec:data_anal}
We obtained the GECAM background-subtracted LCs with 5~ms bin time in the following energy passbands: 30--70, 70--100, 100--150, 150--200, 200--300, 300--500, 500--1000, 1000--6000~keV, respectively. Background interpolation and subtraction were obtained as described in \citet{ZhangWL25}.
In each of these profiles, peaks were identified by means of {\sc mepsa}, a flexible code that was specifically designed to identify peaks in GRB LCs across different timescales \citep{Guidorzi15a}. 
In this present study,  we used both {\sc mepsa} and a faster version of it, which significantly reduces the computing time.\footnote{It is $\sim 100$ times faster than classical {\sc mepsa} at the cost of losing $\sim 5\%$ of all the detected peaks. This is made possible by a sparser sampling of the rebinning factors and related phases. See Maistrello et al. (in prep., submitted to \textit{Astronomy and Computing}) for details.}

Peaks were selected imposing a threshold of S/N~$\ge5$ on the signal-to-noise ratio calculated by {\sc mepsa}.
Each peak is automatically characterised by the following pieces of information: peak time $t_p$ and its uncertainty, which corresponds to the {\sc mepsa} detection timescale $\delta t_p$; peak amplitude $A$ and its uncertainty; estimate of the full width at half maximum (FWHM) of the peak; S/N of the total net counts ascribed to the peak. The FWHM is estimated from the combination of {\sc mepsa} parameters as prescribed in eq.~(A.3) of \citet{Camisasca23}. We studied the evolution with time of peak times, FWHM, and of the waiting times (WT) $\Delta t$, defined as the time intervals between adjacent peaks. We also examined the evolution of $E_{\rm p}$, which is the peak of the $\nu\,F_\nu$ energy spectrum: its values were taken from \citet{Moradi24}.

Hereafter, we focus on the 100--150~keV band LC, which displays the largest number of peaks (103).

\begin{figure}
	\centering 
	\includegraphics[width=0.45\textwidth]{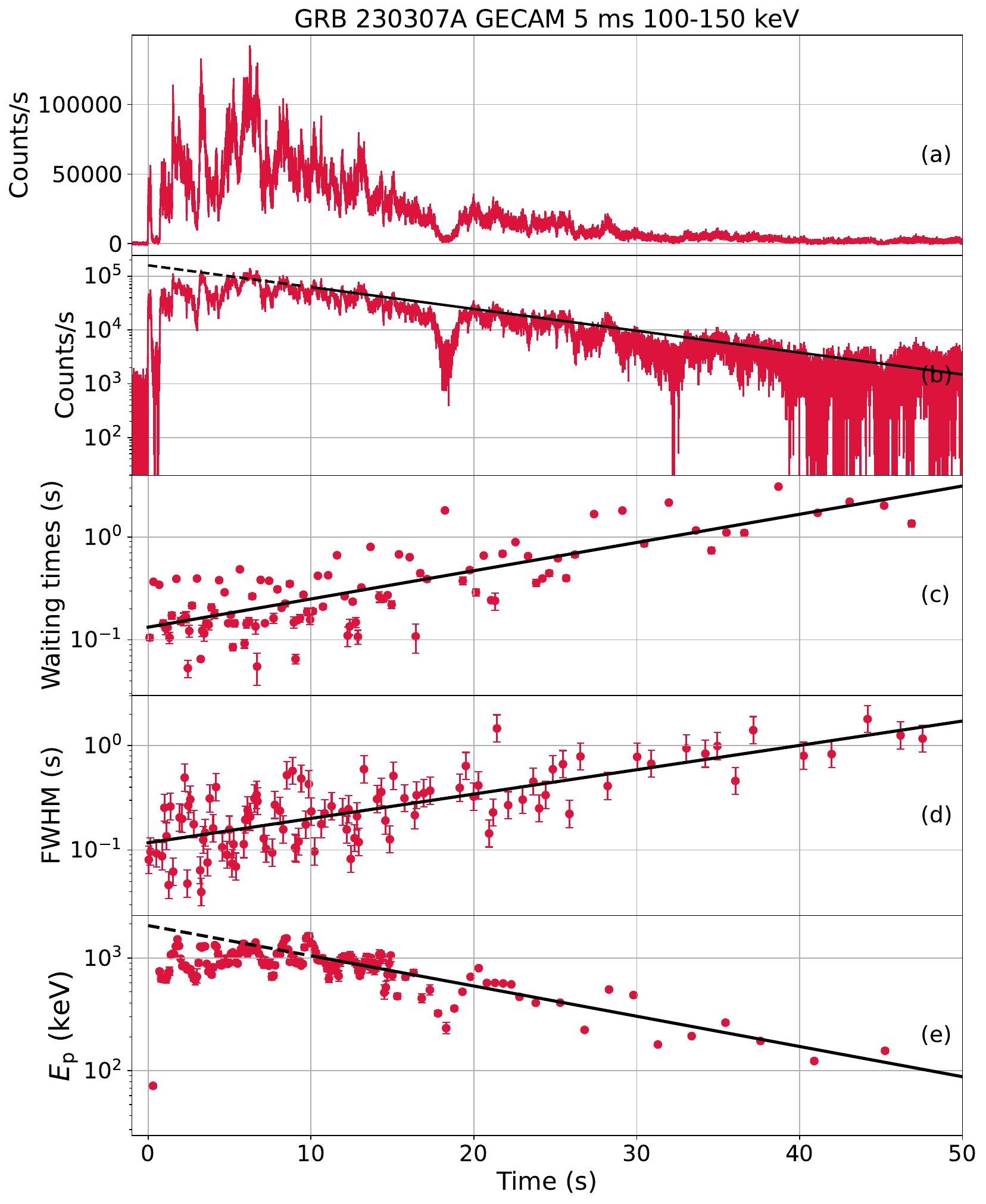}	
	\caption{Properties of GRB\,230307A. (a) LC in the 100-150~keV band with 5-ms bin time. (b) same as (a), but in semi-logarithmic scale;  (c) WTs of the peaks detected with {\sc mepsa} as a function of time; (d) FWHMs of the same peaks of (c) as a function of time; (e) time-resolved evolution of spectral peak energy. In panel (b), the solid line represents the exponential fit of the peak rates of the pulses. In panels (c), (d), and (e), solid lines show the exponential models obtained within the corresponding temporal window, whereas the dashed line, if present, shows the extrapolated fit to the time interval that was ignored by the fitting procedure.} 
	\label{fig:multi}%
\end{figure}

\section{Results and interpretations}
\label{sec:res}
The analysis of the 100-150~keV profile revealed the following properties:
\begin{enumerate}
\item the overall envelope of the LC looks like a so-called FRED. In particular, from $t>10$~s the count rate decays exponentially, covering two decades, as shown in panels (a) and (b) of Fig.~\ref{fig:multi}. Similarly, the peak rates of the pulses detected with {\sc mepsa} also evolve exponentially with time.
Their evolution is described by Eq.~\eqref{eq:pr}:
\begin{equation}
\label{eq:pr}
    P(t) \simeq\ P_0 \, e^{-t/\tau_p}\;,
\end{equation}
with $P_0 = (1.64 \pm 0.12) \times 10^{5}~\rm{cts~s^{-1}} $ and $\tau_p = 10.7\pm 0.4$~s. This model is shown in panel (b) of Fig.~\ref{fig:multi} along with the LC.

\item WTs evolve exponentially with time, spanning two decades, from $\sim 0.1$ to $\sim 10$~s, during the first 50 seconds, as described by Eq.~\eqref{eq:wt},
\begin{equation}
\label{eq:wt}
    \Delta t \simeq\ \Delta t_0\, e^{t/\tau_{\Delta t}}\;,
\end{equation}
with $\Delta t_0 = 0.14\pm 0.02$~s and $\tau_{\Delta t}\simeq 15.8$~s (panel (c) of Fig~.\ref{fig:multi}).

\item The FWHM of the pulses detected by {\sc mepsa} also increases with time: it could be either exponential or linear, spanning more than one decade, from $0.1$~s. If modelled with an exponential, this is described by Eq.~\eqref{eq:FWHM},
\begin{equation}
\label{eq:FWHM}
    {\rm FWHM} (t) \simeq\ {\rm FWHM}_0\  e^{t/\tau_F}\;,
\end{equation}
with FWHM$_0 = 0.12\pm 0.01$~s and $\tau_F = 18.6\pm 1.7$~s. Panel~(d) of Figure~\ref{fig:multi} shows the FWHM values as a function of time. 

We also performed a linear fit FWHM~$= {\rm FWHM}'_{0} + \alpha t$ (see Fig.~\ref{fig:fwhm_lin}) and obtained FWHM$'_{0} =  0.05 \pm 0.01~{\rm s}$ and $\alpha = 0.010 \pm 0.001$, but the result appears to be worse than the exponential model.
We fitted both linear and exponential models, $y=mx+q$, modelling the dispersion as a further parameter, by adopting the D'Agostini likelihood \citep{DAgostini05}. When we used $y = \rm FWHM$, the best-fit parameters are $m = 0.012\pm 0.002$, $q=0.077^{+0.026}_{-0.024}$~s, and $\sigma=0.068^{+0.020}_{-0.017}$~s. Instead, when we use $y = \ln{({\rm FWHM/s})}$, we find
$m = 0.054\pm 0.008$, $q = -2.14\pm 0.14$, and
$\sigma=0.48^{+0.08}_{-0.07}$. The corresponding goodness of the fit was evaluated as $\chi^2 =\sum{ (y_i-y_{{\rm model},i})^2/\sigma_{{\rm tot},i}^2}$, where
$\sigma_{{\rm tot},i}^2 = \sigma^2+\sigma_{y,i}^2 +m^2\sigma_{x,i}^2$, with $\sigma_{y,i}$ and $\sigma_{x,i}$ being the measurement uncertainties of the generic $i$-th point. We obtained
$\chi_{\rm lin}^2 = 134$ and $\chi_{\rm exp}^2=99.6$ (100 degrees of freedom).
\begin{figure}[h!]
    \centering
    \includegraphics[width=0.75\linewidth]{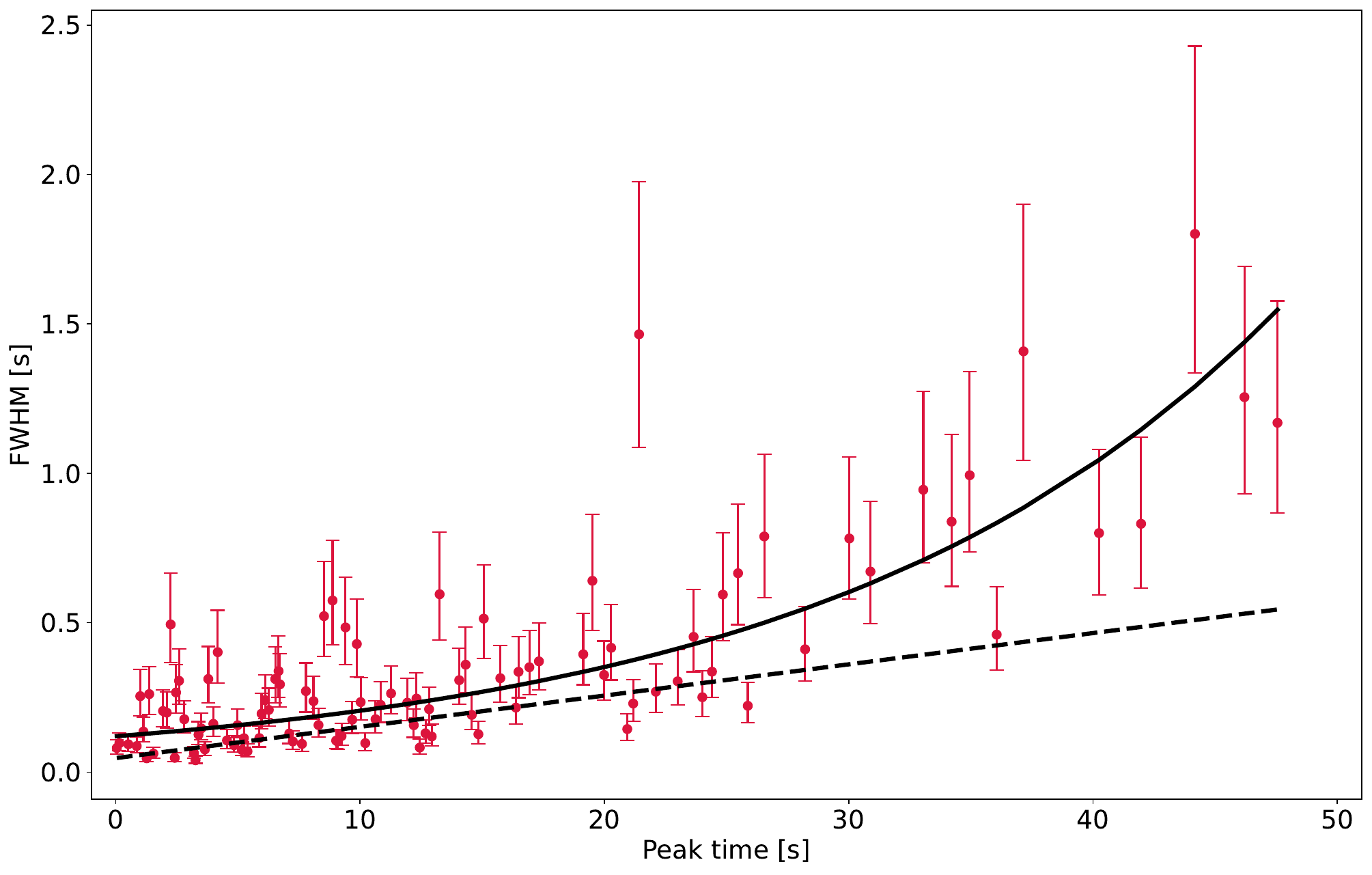}
    \caption{FWHM evolution over time. Solid (dashed) line represents the exponential (linear) model.}
    \label{fig:fwhm_lin}
\end{figure}

\item The evolution of the peak energy, $E_p$, from $t>10$~s can also be described as a negative exponential,
\begin{equation}
\label{eq:Ep}
    E_p(t) \simeq\ (1940\pm 150\ {\rm keV})\  e^{-t/\tau_E}\;,
\end{equation}
with $\tau_E = 16.2 \pm 0.9$~s (see panel~(e) of Fig.~\ref{fig:multi}).
\end{enumerate}

Equations ~\eqref{eq:pr}--\eqref{eq:Ep} were fitted using a nonlinear least squares algorithm (applied on the logarithmic quantities), accounting for y-axis errors. Parameter uncertainties were estimated as the square root of the covariance matrix diagonal. 

\subsection{A simple toy model}
\label{sec:toy_model}
Typically, long and multi-peaked GRB LCs do not show any specific evolution of WTs: moreover, this lack of systematic evolution represents one of the pillars of the IS model as opposed to the ES model as a possible explanation of GRB prompt emission \citep{Fenimore99}.
In this respect, the exponential evolution of WTs exhibited by GRB\,230307A is enough to make it stand out from the population of long and multi-peaked GRBs. Hence, inspired by the unusual WT evolution, we conceived a toy model that can naturally account for it, along with the other observed properties described in Section~\ref{sec:res}.

The sequence of peak times $t_{p,i}$ ($i=1,\ldots,N_p$), where $N_p$ is the total number of peaks, can be described in terms of a stochastic point process: each peak time marks the  occurrence of an event.
In this toy model, there are initially $N_0$ elementary bunches of energy, which share the same probability of ``decaying'' (=releasing their amount of energy within a single shot that manifests itself as a pulse) as well as the same amount of (bolometric) energy. They
``decay'' or release their energy independently of one another. From these simple assumptions, which are the same that rule the radioactive decay of a bunch of $N_0$ atoms of a given element, the exponential distribution of the peak (or release) times $\{t_{p,i}\}$ follows as a consequence. In particular, the mean number of energy bunches still available at time $t$, $N(t)$, is simply given by:
\begin{equation}
\label{eq:Nmodel}
    N(t)\ =\ N_0\ e^{-t/\tau}\;,
\end{equation}
where $\tau$ is the mean lifetime of each bunch. This model predicts the temporal evolution of the expected (or mean) WT, as
\begin{equation}
\label{eq:WT_model}
    \langle \Delta t\rangle\ =\ -\frac{1}{\dot{N}(t)}\ =\ \frac{\tau}{N_0}\, e^{t/\tau}\;,
\end{equation}
which accounts for the observed property modelled in Eq.~\eqref{eq:wt}.

The toy model aims to reproduce a surrogate version of the LC to be compared with that of GRB\,230307A. The key idea is that the LC is the result of the superposition of elementary pulses, whose properties are governed by a few assumptions, which are hereafter defined along with the corresponding model parameters:
\begin{itemize}
    \item $N_0$ peak times are sampled from an exponential distribution with e-folding time $\tau$, in agreement with Eq.~\eqref{eq:Nmodel};
    \item pulse FWHMs are calculated assuming Eq.~\eqref{eq:FWHM}, where FWHM$_0$ and $\tau_F$ are treated as free parameters;
    \item the counts of each pulse is constant and is parametrised by $N_{\rm{cts}}$;
    \item the peak rate $P$ of a given pulse is calculated by dividing the counts by the corresponding FWHM: this choice is corroborated by the fact that $\tau_p$ of Eq.~\eqref{eq:pr} is not wildly different from $\tau_F$ of Eq.~\eqref{eq:FWHM}.
    By doing so, we are implicitly assuming that the peak rate decay and the FWHM rise evolve on a common timescale, $\tau_F$. This assumption is not driven by any specific theoretical interpretation but rather stems from the observation of the data and adopting Occam’s razor.
    To model the fact that the overall LC shows a finite rise time, we added an extra term that accounts for the suppressed flux of the initial pulses over a timescale described by the parameter $\tau_r$, which acts as the rise time. As a result, peak rates are calculated as a function of the corresponding peak times $t_p$ as
    \begin{equation}
    \label{eq:prnew}
    P(t_p) = \frac{N_{\rm{cts}}}{{\rm FWHM}_0}\ e^{-t_p/\tau_F}\ (1-e^{-t_p/\tau_{r}})\;.
    \end{equation}
    The corresponding additional free parameter is $\tau_r$.
    \item Concerning the pulse shape, with reference to the \citet{Norris96} model, the peakedness is fixed to $\nu =2$, while the decay-to-rise time ratio $r$ to 3 for all pulses. Since the FWHM is calculated for each pulse, rise and decay times are consequently computed as $t_r = {\rm FWHM}/((1+r)(\ln{2})^{1/\nu}) \simeq {\rm FWHM}/3.33$ and $t_d = r t_r$, respectively.
\end{itemize}
%

\subsection{Physical model 1: multiple shells emitted with equal Lorentz factors}
\label{sec:phys_model1}
In the toy model of Section~\ref{sec:toy_model}, the FWHM is assumed to evolve exponentially. Here, we adopted another FWHM evolution, derived from shock kinematics. In this model, the central engine is working in the same way as in Section~\ref{sec:toy_model}, but, in addition, the train of shells emitted by the central engine is colliding with a so-called target shell. The target shell is expanding, so successive shocks occur at increasing radii. The target shell could have been expelled some time before the train of shells or right at the beginning of the merger. Alternatively to the target shell, the various shells might be refreshing the external blast wave, as it was suggested in the refreshed shock scenario to explain long-lived afterglows \citep{Rees98}.The origin of the target shell is further discussed in Section~\ref{sec:disc}.

To be more quantitative, we let $t_{e,i}$ be the emission time (lab frame) of the $i$-th shell. All shells are emitted with the same velocity $\beta=v/c$, whereas the target shell is moving with $\beta_s = v_s/c < \beta$. We let $R_0$ be the initial radius of the target shell at $t=0$, when the first shell is emitted. We define $t_{c,i}$ the collision time (lab frame) of the $i$-th shell, which takes place at radius $R_{c,i}$. Since different collisions take place at different radii, the WTs measured by the observer will appear shorter than the corresponding WTs in the lab frame: to avoid confusion, the observed collision times are denoted as $t_{c,i}^{\rm (obs)}$ and are measured by the observer since the arrival of the first pulse. The kinematic solution for these quantities is
\begin{equation}
    t_{c,i}\ =\ \frac{R_0}{v - v_s} + \Big(\frac{v}{v - v_s}\Big)\,t_{e,i}\;,
    \label{eq:tci}
\end{equation}
\begin{equation}
    t_{c,i}^{\rm (obs)}\ =\ \Big(\frac{\beta}{\beta - \beta_s}\Big)\,(1 - \beta_s)\,t_{e,i}\;,
    \label{eq:tciobs}
\end{equation}
\begin{equation}
    R_{c,i}\ =\ \Big(\frac{\beta}{\beta - \beta_s}\Big)\,(R_0 +  v_s\,t_{e,i})\;.
    \label{eq:Rci}
\end{equation}
From Eq.~\eqref{eq:tciobs} the observed WTs, $\Delta t_{c,i}^{\rm (obs)}$, inherit the same exponential temporal evolution of the lab-frame WTs, $\Delta t_{e,i}$. The duration of each pulse is given by the angular timescale: because of the progressive expansion of the target shell, durations are predicted to increase linearly with time,
\begin{equation}
    {\rm FWHM}_i\ =\ \frac{R_{c,i}}{2 c \Gamma^2}\ =\ \frac{1}{2 c \Gamma^2}\ \Big(\frac{\beta}{\beta - \beta_s}\Big)\,(R_0 +  v_s\,t_{e,i})\;.
    \label{eq:toyFWHM}
\end{equation}

That the velocity of the target shell would change only negligibly could be explained by its being more massive than the faster shells. If both fast and target shells are relativistic, the term $\beta/(\beta - \beta_s)$ is approximately $2\Gamma^2/((\Gamma/\Gamma_s)^2 - 1)$. In particular, Eqs.~\eqref{eq:tciobs} and \eqref{eq:toyFWHM} would become,
\begin{equation}
    t_{c,i}^{\rm (obs)}\ \simeq\ \frac{1}{1 - (\Gamma_s/\Gamma)^2}\ t_{e,i}\;,
    \label{eq:tciobs_approx}
\end{equation}
\begin{equation}
    {\rm FWHM}_i\ \simeq\ \frac{R_0/c + \beta_s\,t_{e,i}}{(\Gamma/\Gamma_s)^2 - 1}\;.
    \label{eq:toyFWHM_approx}
\end{equation}

Within this interpretation, from Eq.~\eqref{eq:toyFWHM}, FWHM is expected to grow linearly with time, in possible agreement with observations. Hence, to simulate the model, we replaced Eq.~\eqref{eq:FWHM} with Eq.~\eqref{eq:toyFWHM}, and modified Eq.~\eqref{eq:prnew} to take into account the linear evolution of the FWHM. Furthermore, the emission times $t_{e,i}$ are still generated from an exponential distribution, but need to be transformed into the observed collision times, using Eq.~\eqref{eq:tciobs}. 

We initially assumed a non-relativistic target shell, $\beta_s<1$. This led to an untenable physical solution, in particular because of the compactness problem\footnote{In this case, we obtained $\beta_s \sim 0.4$, $R_0 = 3 \times 10^{10}~{\rm cm}$, and $\Gamma \sim 6.5$, implying sub-photospheric shocks between $5 \times10^{10}$ and $10^{12}~{\rm cm}$, with pair-production opacity $\tau_{\gamma \gamma}$ (computed in Sect.~\ref{sec:test}) remaining extremely high (from $10^9$ initially to $10^7$)} throughout the burst.
(see Sect.~\ref{sec:test}). We consequently adopted $\Gamma_s$ instead of $\beta_s$ as a more convenient parameter, implicitly assuming an ultra-relativistic motion for the target shell.

\subsection{Physical model 2: multiple shells emitted with gradually decreasing Lorentz factors}
\label{sec:phys_model2}
This model is almost the same as the previous one described in Section~\ref{sec:phys_model1}, except for one assumption: the Lorentz factors of the different shells decrease with the emission times, so that later shells are slower.
In this model, the previous equations Eqs.~\eqref{eq:tci}--\eqref{eq:toyFWHM} are still valid, but since the Lorentz factor is changing from one emitted shell to another, $\Gamma$ has to be replaced by $ \Gamma_i \equiv\Gamma(t_{e,i})$. Prompted by the exponential evolution of FWHM and peak energy (Eqs.~\ref{eq:FWHM} and \ref{eq:Ep}), we parametrised the temporal evolution of $\Gamma(t)$ as 
\begin{equation}
\label{eq:gamma}
\Gamma(t) = \Gamma_0e^{-t/\tau_{\Gamma}}.
\end{equation}
Instead of $\Gamma$ of the model of Sect.~\ref{sec:phys_model1}, this scenario requires a couple of new parameters, $\Gamma_0$ and $\tau_{\Gamma}$, through which each $\Gamma_i$ is calculated.

\subsection{Genetic algorithm based parameter optimisation}
\label{sec:ga}
We determined the free parameters of each model using a genetic algorithm (GA; see \citealt{Bazzanini24,Maistrello25} for similar usages). The loss function to be minimised by the GA consists of four different contributions. Each term aims to ensure that the corresponding property of the real LC of GRB\,230307A is correctly reproduced by the simulated LC. The four properties include:
\begin{enumerate}
    \item similar envelopes (smoothed time profiles);
    \item compatible distributions of peak times;
    \item comparable numbers of peaks detected with {\sc mepsa};
    \item compatible distributions of FWHM values.
\end{enumerate}
A detailed description of the GA-based optimisation procedure is reported in \ref{sec:model_optim}. 

\subsubsection{Results for the toy model}
\label{sec:res_model1}
The best model parameters that we have come up with for the toy model of Sect.~\ref{sec:toy_model} are shown in Table~\ref{tab:model_comparison}.
We converted the pulse counts $N_{\rm cts}$ to fluence $N_{F}$ in erg~cm$^{-2}$. This was done by multiplying $N_{\rm cts}$ by the fluence-to-counts ratio $R$ computed using the values of the time-resolved spectral modelling of \citet{Moradi24}. For a constant $N_{\rm cts}$, $N_F$ decreases with time; thus, we adopted an average value of $1.8 \times 10^{-7}$~erg~cm$^{-2}$ in the 100-150 keV energy range. After applying the $k$-correction, the fluence of a single shot over the full energy band (6-8000 keV) is $3.9 \times 10^{-6}$~erg~cm$^{-2}$, which, placed at a distance of $300~{\rm Mpc}$ ($z=0.065$), corresponds to an isotropic-equivalent energy of $\sim 4.1~\times~10^{49}$~erg. Given that the total time-integrated fluence is $4.8 \times 10^{-3}$~erg~cm$^{-2}$, the fluence of a single elementary pulse, multiplied by the number of pulses, roughly matches the total GRB fluence. The central engine is emitting numerous energy bunches with an average energy of $4.1\times 10^{49}$~erg. Earlier bunches are more energetic, ranging from $2.5 \times 10^{50}$ all the way down to $1\times10^{48}$ erg. 

Noticeably, the intrinsic number of bunches of energy, $N_0$, which is also the number of intrinsic peaks that make up the overall profile, is about ten times higher than the number of {\sc mepsa}-detected peaks (1000 vs 100). This result suggests that the observed LC consists of a myriad of short overlapping peaks, which blend together and appear as fewer, broader peaks.
\begin{figure}
    \centering
    \includegraphics[width=0.45\textwidth]{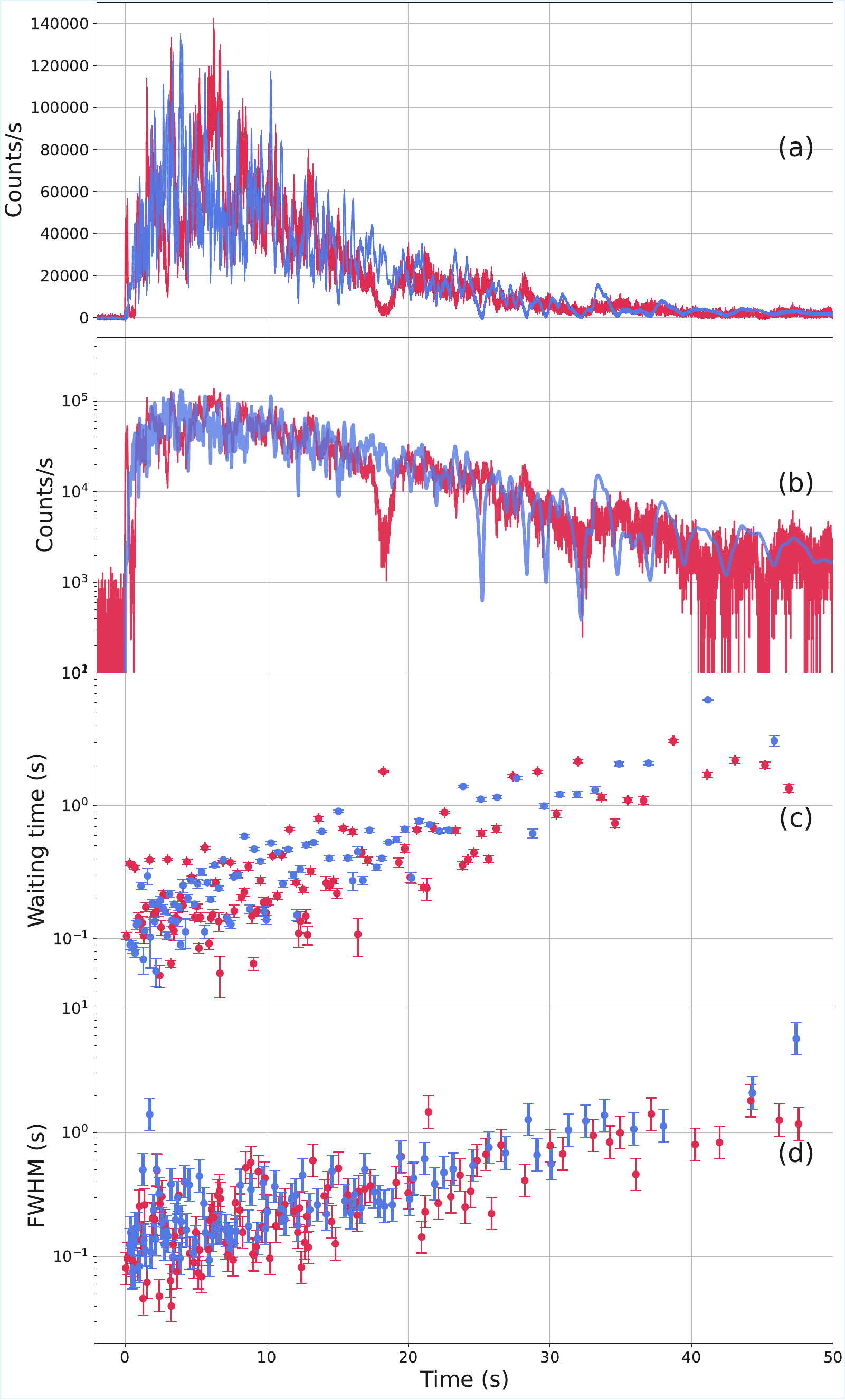}
    \caption{Panels (a) and (b) illustrate the temporal evolution of count rates, (a) in linear and (b) in logarithmic scale. Panels (c) and (d) show the temporal evolution of the WTs and of the pulse FWHMs, respectively. Red points show real data, while blue points are simulated according to the toy model of Section~\ref{sec:toy_model}.}
    \label{fig:multi_toy}
\end{figure}

We generated a synthetic profile and compared it and its properties with the corresponding real ones: Figure~\ref{fig:multi_toy} shows the overall LC in both linear (a) and logarithmic scale (b). The temporal evolution of WTs and of FWHMs of the peaks detected with {\sc mepsa} are displayed in panels (c) and (d), respectively. All the temporal properties that are found in the real data are faithfully reproduced by the synthetic profile.
A possible weakness of this model is that the broad dip observed at 20 s can hardly be reproduced in its depth.
\subsubsection{Results for the physical models}
\label{sec:res_phys_models}
\begin{figure*}
    \centering
    \includegraphics[width=0.45\textwidth]{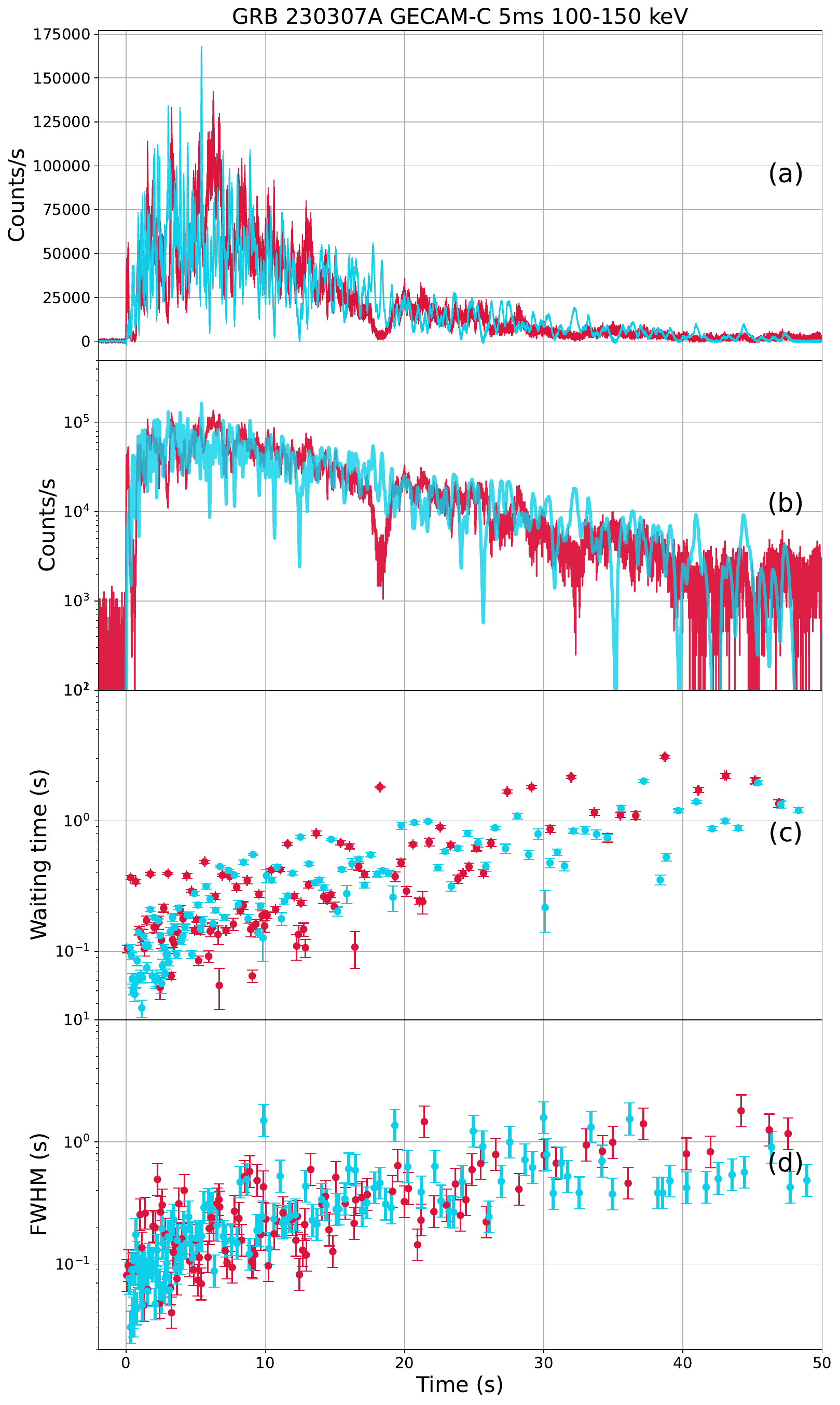}
    \includegraphics[width=0.45\textwidth]{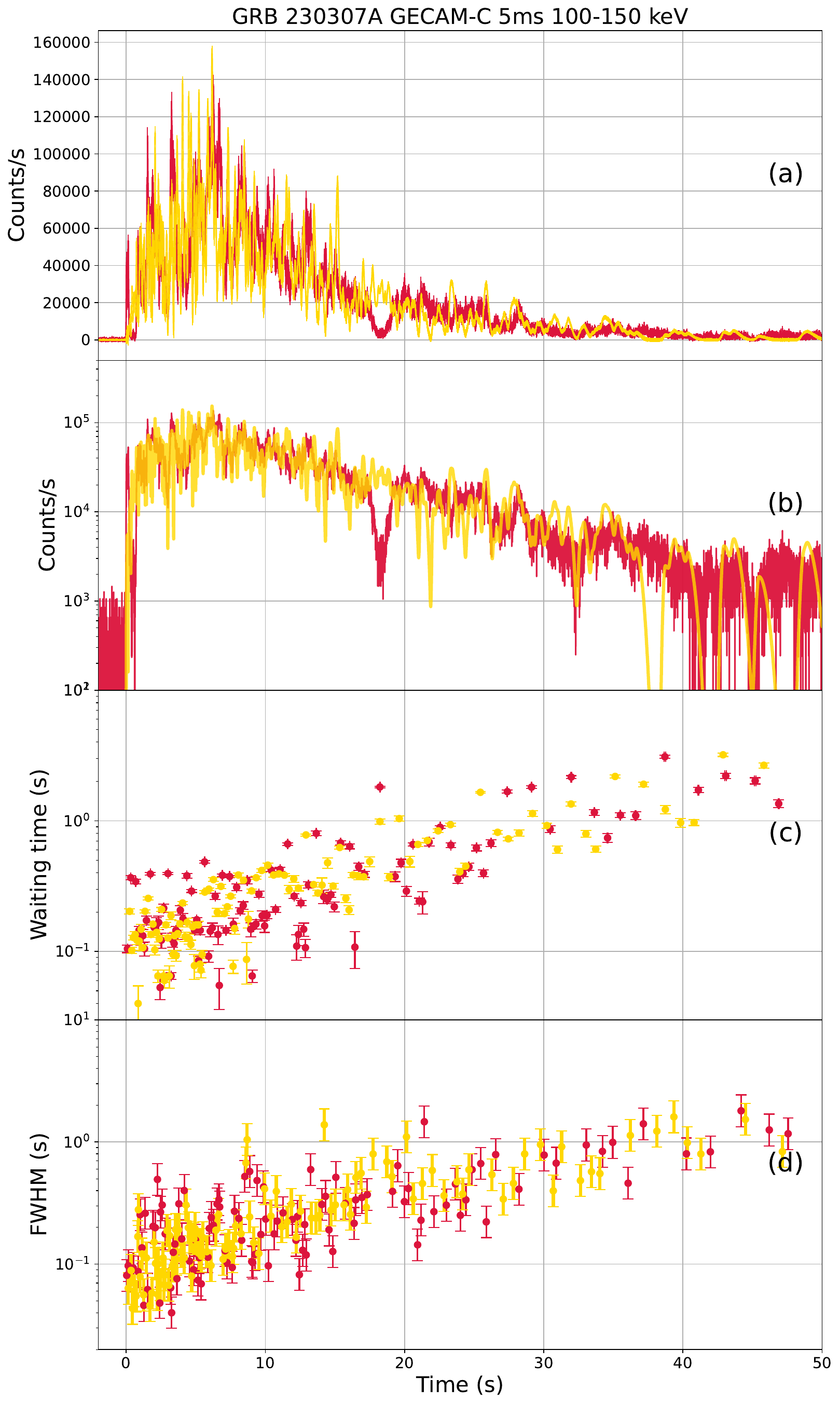}
    \caption{Same as in Figure~\ref{fig:multi_toy}, for the physical model with a wind of shells with constant $\Gamma$ described in Section~\ref{sec:phys_model1} (left panel) and the one with declining $\Gamma$'s described in Section~\ref{sec:phys_model2} (right panel). Real data are shown in red, while simulated data from constant and declining $\Gamma$ model are shown in cyan and yellow, respectively.}
    \label{fig:multi_toy_new}
\end{figure*}

The results for the constant-$\Gamma$ scenario (Section~\ref{sec:phys_model1}) and for the declining-$\Gamma$ factor scenario (Section~\ref{sec:phys_model2}) are shown in Table~\ref{tab:model_comparison}.
\begin{table*}[ht]
\centering
\small 
\setlength{\tabcolsep}{4pt} 
\begin{tabular}{@{}l@{\hskip 6pt}c@{\hskip 6pt}c@{\hskip 6pt}c@{}}
\toprule
Parameter & Toy Model & Physical Model 1 (Constant $\Gamma$) & Physical Model 2 (Declining $\Gamma$) \\
\midrule
\midrule
$N_0$ & $862^{+1}_{-25}$ & $1270^{+110}_{-90}$ & $894^{+21}_{-101}$ \\
$\tau$ [s] & $10.9^{+0.1}_{-0.5}$ & $10.9^{+1.0}_{-2.3}$ & $10.1^{+1.4}_{-0.8}$ \\
FWHM$_0$ [s] & $0.05^{+0.01}_{-0.01}$ & — & — \\
$\tau_F$ [s] & $14.0^{+0.2}_{-0.4}$ & — & — \\
$\tau_r$ [s] & $2.5^{+0.1}_{-0.7}$ & $2.9^{+0.3}_{-0.5}$ & $3.4^{+0.7}_{-0.9}$ \\
$N_{\text{cts}}$ & $1449^{+26}_{-5}$ & $920^{+30}_{-20}$ & $1431^{+49}_{-20}$ \\
$\Gamma$ or $\Gamma_0$ & — & $94^{+8}_{-1}$ & $339^{+18}_{-20}$ \\
$\Gamma_s$ & — & $8.6^{+0.1}_{-0.4}$ & $26.2^{+0.5}_{-0.5}$ \\
$\tau_\Gamma$ [s] & — & — & $96^{+30}_{-17}$ \\
$R_0$ [cm] & — & $9.1^{+3.8}_{-1.5} \times 10^{10}$ & $1.4^{+0.4}_{-0.3} \times 10^{11}$ \\
\midrule
\bottomrule
\end{tabular}
\caption{Parameters of the three GRB emission models used to reproduce the temporal evolution of GRB\,230307A.}
\label{tab:model_comparison}
\end{table*}
Figure~\ref{fig:multi_toy_new} shows the corresponding results analogously to Figure~\ref{fig:multi_toy}.
In the last scenario considered, the results show that $\Gamma$ decreases relatively slowly for the model to reproduce the observed GRB time profile. Assuming a constant fluid-comoving-frame peak energy—thus attributing all observed peak energy variation to a decreasing $\Gamma$—is ruled out. In fact, a decay timescale of $\tau_{\Gamma} \sim 100~{\rm s} $ is not short enough to explain the drop seen in Panel (e) of Figure~\ref{fig:multi}. The observed peak energy decline should be a consequence of the decreasing shock energy density resulting from the expansion of the emission radius.

\subsection{Testing the physical constraints: compactness problem, photosphere, and internal shock radii.}
\label{sec:test}
%
We verified that the compactness problem is not an issue for our model. To this aim, we have computed the optical depth to pair production \begin{equation}
\label{eq:optical_depth_pair}
    \tau_{\gamma\gamma} = \frac{f_p\sigma_T \phi D_L^2 }{\Gamma^3 R_e^2m_ec^2},
\end{equation} with $f_p$ the fraction of photons making pairs, $\sigma_T$ the Thomson scattering cross section, $\phi$ the fluence, $D_L$ the luminosity distance, $R_e$ the emission radius, $m_e$ the electron mass, and $c$ the speed of light \citep{Piran04_rev}. We evaluated $\tau_{\gamma\gamma}$ at early and late times, taking for $R_e$ the radius of the first and last shocks, respectively. Given that the first pulses produced by the model have a width of about $25~{\rm ms}$, from Eq.~\eqref{eq:toyFWHM} the initial shock radius is about $1.3 \times 10^{13}~{\rm cm}$ and $\tau_{\gamma\gamma} \simeq 19\times f_p$, while for the last wider pulses of about $1~{\rm s}$, the opacity drops to $\tau_{\gamma\gamma} = 1.2\times10^{-2} \times f_p $. The high initial opacity may account for the early rise in the observed flux, which in our model is modeled by the rise time $\tau_r$. Consequently, the rising phase could reflect the transition from an initially optically thick to an optically thin regime as the emitting region expands.

We also computed the photospheric radius, $R_{\rm ph}$, to evaluate whether any emission in our model could originate below the photosphere, which would typically give rise to a thermal spectral component not observed in the data
\begin{equation}
\label{eq:photospheric_radius}
    R_{\rm ph} = \frac{L\sigma_T}{8\pi m_p c^3 \Gamma^3},
\end{equation}
here $L$ is the isotropic luminosity, $m_p$ the proton mass \citep{Daigne02}. Using $L= 7.6 \times 10^{51}~{\rm erg~s^{-1}}$ \citep{Svinkin23_230307A}, it is $R_{\rm ph} = 5.5 \times 10^{12}~{\rm cm}$, which is lower than the radius of the first shocks ($1.3 \times 10^{13}~{\rm cm}$). Hence, all shocks in our model occur above the photosphere, consistent with the absence of a detectable thermal component in the prompt emission spectrum.

Finally, we verified that internal shocks between two fast shells do not occur before the collision with the target shell. While some variability in the ejection velocities is expected and may lead to internal shocks among the fast shells themselves, we computed the corresponding internal shock radius using \begin{equation}
    R_{\rm IS} = \Gamma^2c\Delta t, 
\end{equation}
where $\Delta t$ represents the time interval between the emission of two consecutive shells \citep{Daigne98}. Assuming a typical value of $\Delta t = 0.1~{\rm s}$ for the early shells, we found $R_{\rm IS} \simeq 2.6 \times 10^{13}~{\rm cm}$, which lies beyond the radius of the first shock with the target shell. Since the WTs increase more rapidly than the pulse widths, $R_{\rm IS}$ continues to grow and remains larger than the radius of the shocks between the fast and target shells. This confirms that the shell collisions between the fast shells and the slow target shell always occur before any possible internal shocks between fast shells.

\subsection{Similar analysis of other known long COM candidates.}
\label{sec:otherCOMs}
\begin{figure}[h!]
    \centering
    \includegraphics[width=0.5\textwidth]{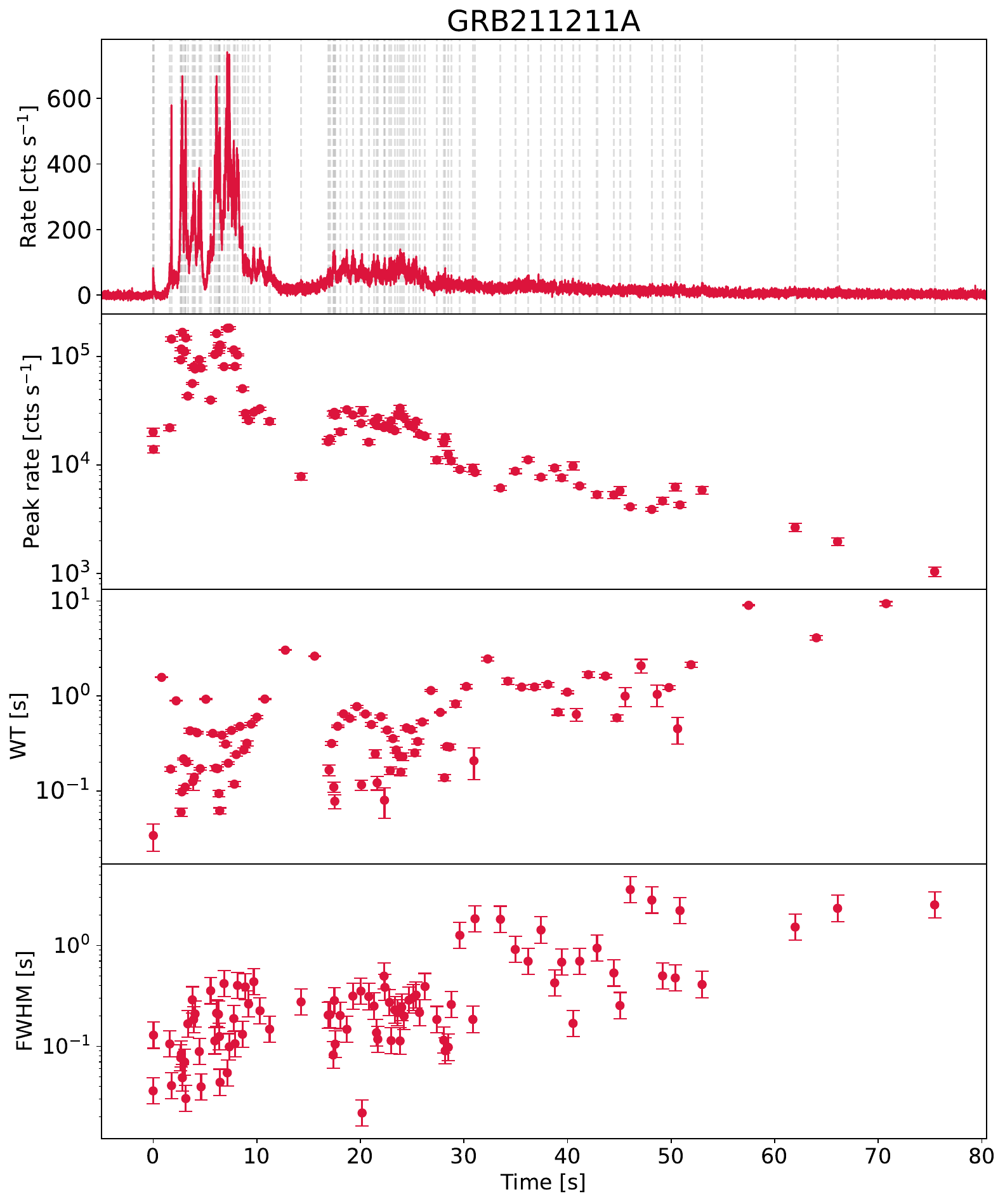}
    \caption{GRB\,211211A \textit{Fermi}/GBM 8-1000 keV LC binned at 4 ms.}
    \label{fig:GRB211211A}
\end{figure}

\begin{figure}[h!]
    \centering
    \includegraphics[width=0.5\textwidth]{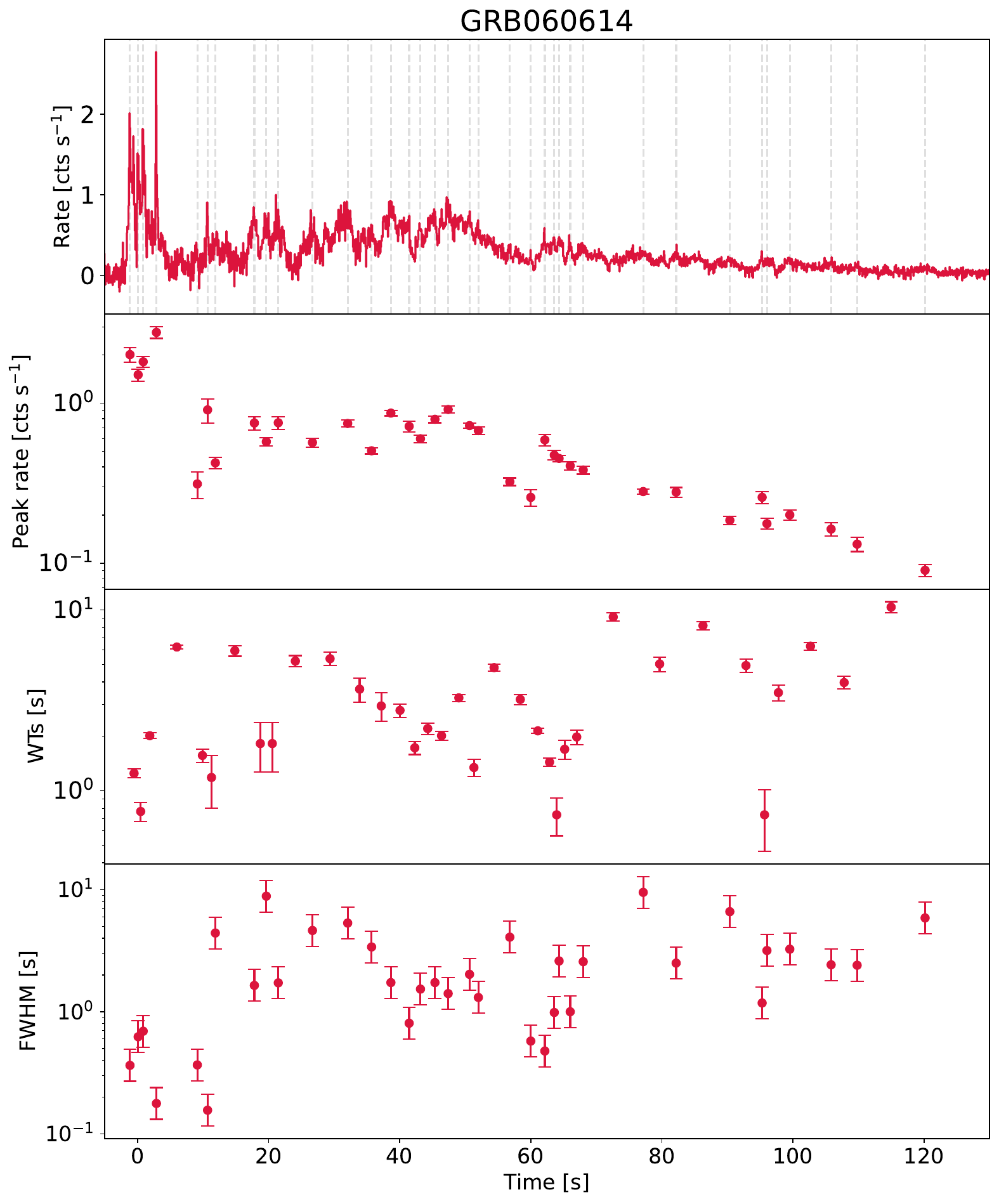}
    \caption{GRB\,060614  \textit{Swift}/BAT 15-350 keV LC binned at 64 ms.}
    \label{fig:GRB060614}
\end{figure}

We carried out a similar and preliminary analysis to other known LGRBs that are COM candidates with enough peaks: GRB\,211211A and GRB\,060614. Figures \ref{fig:GRB211211A} and \ref{fig:GRB060614} illustrate that these events exhibit similar trends as those observed in GRB\,230307A, suggesting that such properties might be characteristic of this subclass of GRBs. A more detailed analysis of their temporal evolution is reported in \ref{sec:other_mergers}. To test whether these properties can be taken as indicators of a COM origin, we applied the same analysis to a LGRB with enough peaks and that is associated with a Ic broad-lined supernova and for which, therefore, a merger origin is excluded with confidence. 

\begin{figure}[h!]
    \centering
    \includegraphics[width=0.5\textwidth]{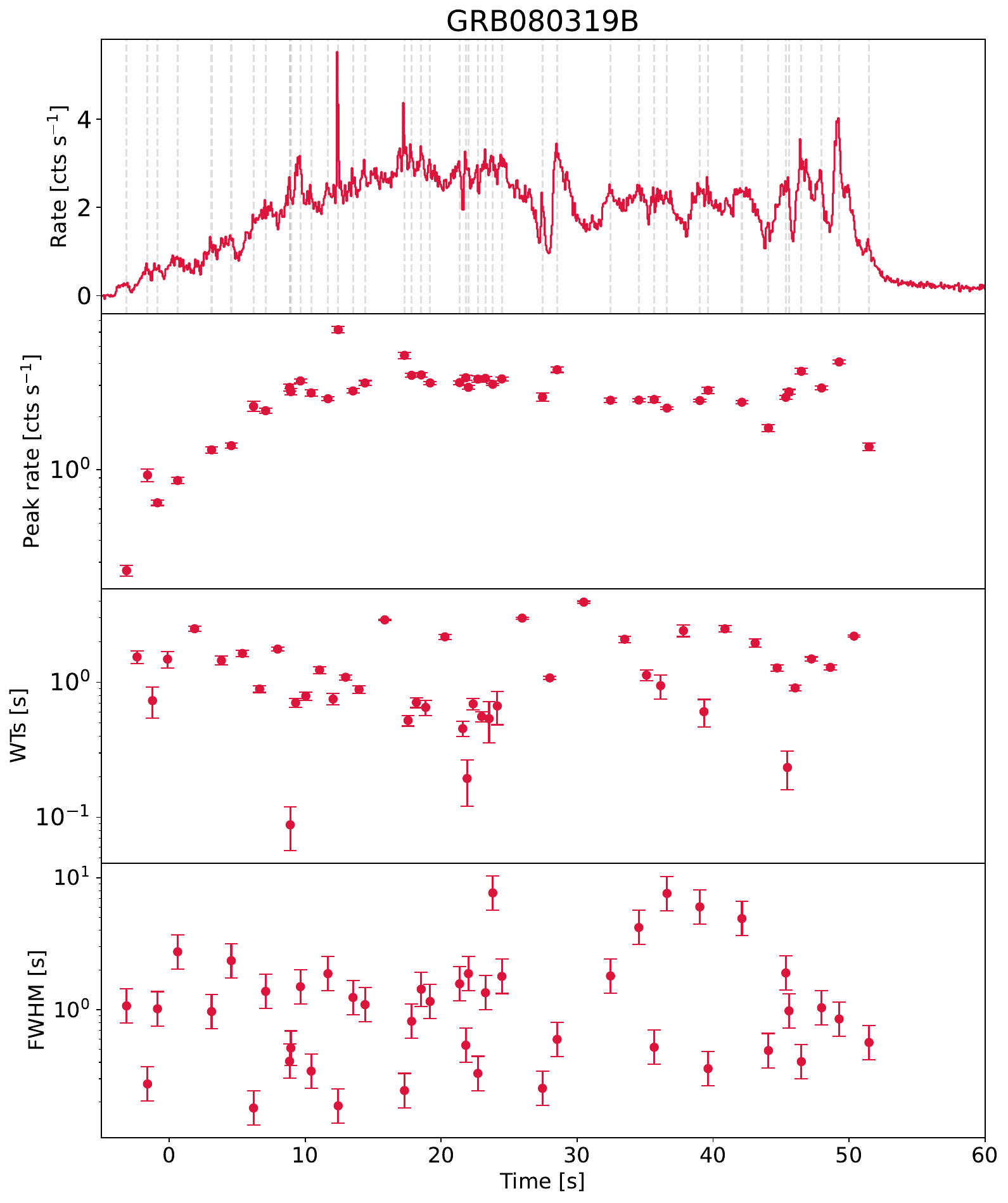}
    \caption{Naked-eye burst GRB\,080319B \textit{Swift}/BAT 15-350 keV LC binned at 64 ms. { Unlike COM candidates GRB\,230307A, GRB\,211211A, and GRB\,060614, this collapsar event does not show the same monotonic evolution of the various observables.}}
    \label{fig:GRB080319B}
\end{figure}

To this aim, the well-known naked-eye burst GRB\,080319B \citep{Racusin08} represents an ideal test-bed: the exceptional quality of its {\em Swift}/BAT 15--150~keV spiky LC offers an excellent opportunity to detect possible temporal trends in WTs and FWHMs. However, as shown in Figure~\ref{fig:GRB080319B}, no such evolution of WTs or of FWHMs with time emerges, in full agreement with an IS interpretation. 
{We extended the same test to the other known SN-associated GRBs having a LC with a suitable number of detected peaks for the same analysis to be applied. To this aim, we could identify six SN-GRBs, in addition to the extraordinary case of GRB\,221009A: their analysis is reported in \ref{sec: collapsar}.
As a result, none of them exhibits the systematic exponential evolution of WT and of FWHM seen in GRB\,230307A (see Fig.~\ref{fig:5SN_GRBs}).}
Further analysis involving a larger sample of GRBs is planned for future work, but the current examples already suggest that such temporal characteristics can serve as valuable indicators to identify long COM candidates from {their} prompt emission profile. 

%
\section{Discussion}
\label{sec:disc}
Contrary to the majority of multi-peaked LGRBs, GRB\,230307A remarkably displays a number of properties that evolve with time exponentially. For other COM candidates, it was already found that both $E_{\rm p}$ and X-ray luminosity appear to decline exponentially with time \citep{Gompertz23}: in this respect, our findings reinforce the role of an exponential evolution for the COM candidates and extend it to other properties, such as peak times, durations, WTs, and peak rates of the pulses that make up the GRB prompt emission.

The deterministic evolution of WTs and of FWHMs with time observed in GRB\,230307A directly clashes with the key idea of the IS model, which was conceived at the time to explain the apparent lack of evolution of timescales within a number of BATSE bursts \citep{Fenimore99}. Rather, the systematic increase of pulse durations with time, as pointed by the FWHM evolution, agrees with the expectations of a simple ES model: as the blast wave expands, the duration of pulses, supposedly caused by interactions with the medium inhomogeneities, increases with time \citep{Fenimore96,Dermer99,Dermer08}.

\subsection{Emission from multiple shells emitted by the central engine}
\label{sec:multiple_shell_toy_model}
Our model assumes that multiple shells (corresponding to the energy bunches described in Section~\ref{sec:toy_model}) are emitted by the central engine and are colliding with either a previously emitted, more massive target shell that is expanding more slowly, or alternatively with the external blast wave as it was suggested to explain long-lived afterglows in the so-called refreshed {shocks} scenario \citep{Rees98}. To account for the observed rapid variability and high-energy emission in the gamma-ray band -- rather than in the optical -- the collisions must occur at relatively small radii ($\sim 10^{13}$ cm), as discussed in Section 3.5. If the target shell corresponds to an external blast wave, the deceleration radius of the jet (or the radius at which a significant portion of  jet begins to decelerate) must lie within the collision radius. This would imply that the merger occurred in an unusually high-density environment, and a bright afterglow would be expected. However, a low circumburst density of approximately $10^{-5}$–$10^{-4}~\mathrm{cm}^{-3}$ is required to explain both the non-detection of GeV emission and the multi-band afterglow data for this event \citep{Dai24}. Therefore, the refreshed shock scenario -- where the target shell is a blast wave -- is disfavoured by these observations. 
    
A plausible candidate for the slowly expanding target shell is the dynamical ejecta produced during the merger (e.g., neutron star–neutron star merger). Tidal interactions and shocks experienced by the neutron stars near the time of coalescence lead to the ejection of material on a dynamical timescale (e.g. \citealt{Dietrich17,Radice18,Shibata19,Rosswog25}). The tidal component of the dynamical ejecta is launched first, predominantly along the orbital plane, followed by a more isotropic, shock-driven ejecta component. Although the bulk of the merger ejecta is sub-relativistic, material along the rotational (jet) axis may be further accelerated by the leading edge or early-ejected component of the relativistic jet. Since the first interaction likely occurs well inside the photosphere, the associated shock radiation would be totally suppressed. Instead, the shock energy is efficiently converted into the bulk kinetic energy of the target shell \citep{Kobayashi02}. Given the low baryon contamination along the jet axis, a significant boost in velocity may be achievable.

Another, more ad hoc possibility relates to the nature of the central engine. Since the jet acceleration process is still poorly understood, it is possible that the engine initially ejects a slower, more massive shell, followed by faster, less massive ejecta. In our model, the initial radius of the target shell is $9 \times 10^{10}~{\rm cm}$. Given its relativistic velocity, this suggests that the massive, slow shell was expelled approximately 3 seconds prior to the rest of the ejecta in this scenario.

The central engine may be powered by a black hole accretion disc, into which fragmented material accretes independently. The accretion of each fragment leads to the ejection of a shell, which subsequently collides with the slower target shell at progressively larger radii. This simple toy model predicts a sequence of pulses that can be described as a piecewise Poisson process, whose shot rate gradually decreases with time (for further details, see \citealt{Guidorzi15b}).
This mechanism explains the exponential distribution of the peak times as well as the exponential evolution with time of their WTs expressed by Eq.~\eqref{eq:WT_model}. In the so-called toy model, the FWHM evolution is phenomenological: to match the observed trend, an exponential evolution seems at first sight the best solution. However, the evolution of the FWHM could be directly inferred from shock kinematics, using Eq.~\eqref{eq:toyFWHM}.

\begin{figure*}
    \centering
    \includegraphics[width=0.5\textwidth]{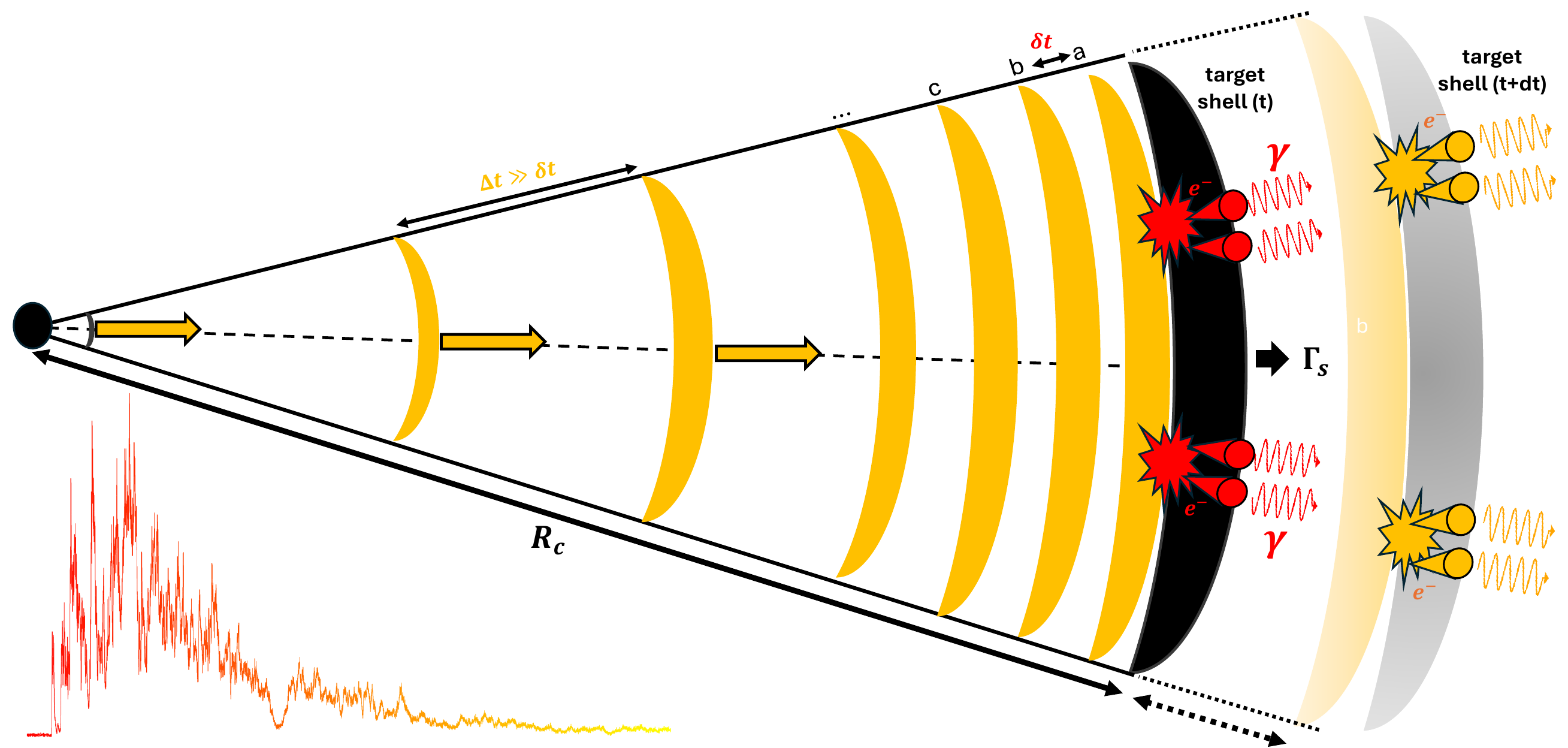}
    \vspace{0.5cm}
    \hrule
    \vspace{0.5cm}
    \includegraphics[width=0.5\textwidth]{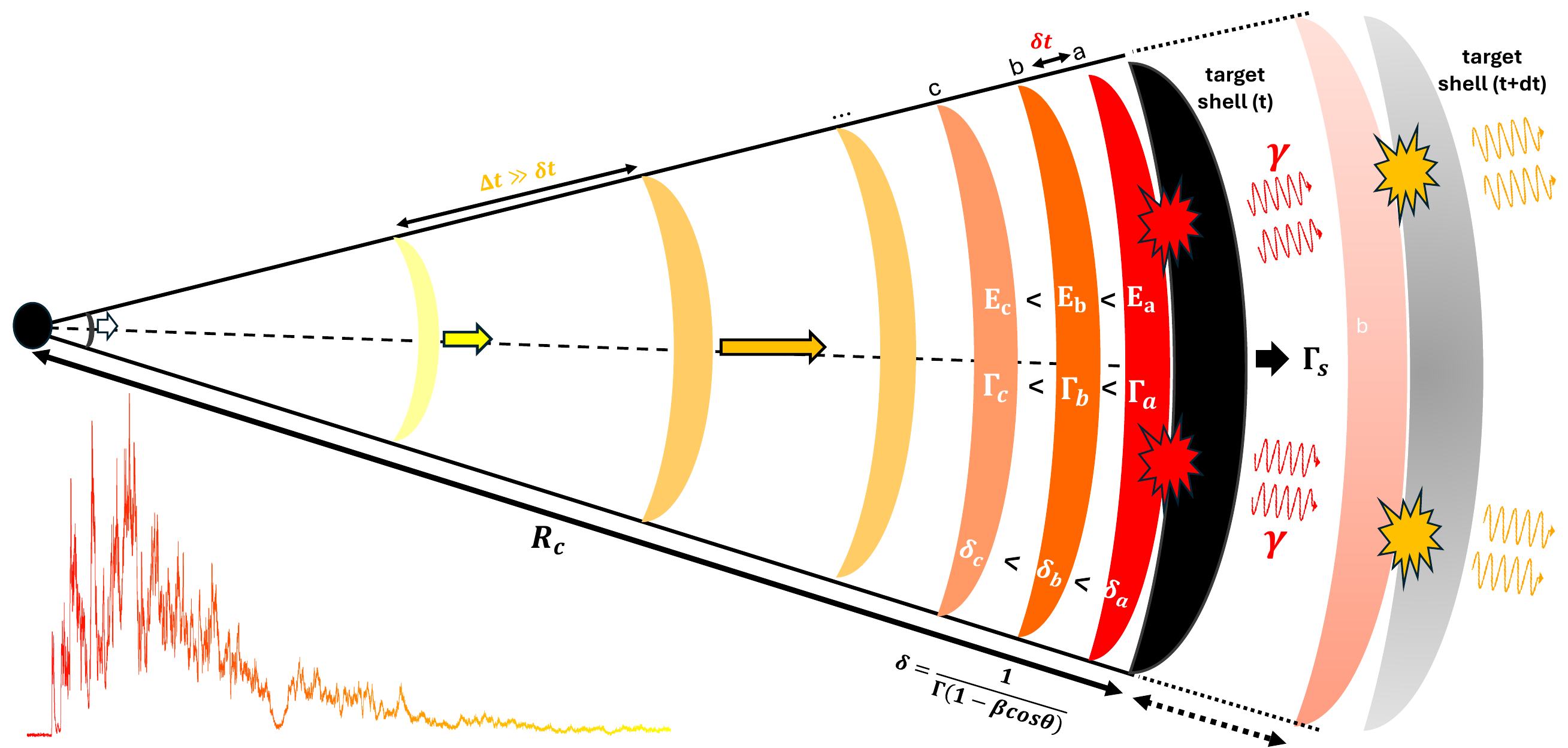}
    \caption{Sketch of the emission from multiple shells in the (i) constant Lorentz factor (Sect.~\ref{sec:phys_model1}) model (top panel) and in  the (ii) decreasing Lorentz factor (Sect.~\ref{sec:phys_model2}) model (bottom panel). Energy bunches emitted from the central engine collide with a slower target shell (shown in black) that was emitted earlier by the central engine. In (i), all shells nearly have the same {Lorentz factor}, while in (ii) the later-emitted shells (in yellow or light orange) are slower and carry less energy than the earlier ones (in deep orange and red). The first shocks correspond to the intense and narrow peaks observed in the early prompt emission of GRB\,230307A, while the later shocks are associated with the dim and broad pulses seen in the extended emission phase.}

\label{fig:toy_model}
\end{figure*}

We considered possible evolutions of the Lorentz factor $\Gamma$ of the various shells: (i) approximately constant (Sect.~\ref{sec:phys_model1}); (ii) gradually decreasing with time (Sect.~\ref{sec:phys_model2}). In scenario (i), the ${\rm FWHM}$ is increasing linearly with time.
Alternately to (i), scenario (ii) assumes a gradually decreasing Lorentz factor: this naturally explains both the evolution of ${\rm FWHM}$, which can also be stronger than linear, as well as a decreasing Doppler boosting. In scenario (ii), the pulses' broadening and the spectral softening share a common explanation. The decaying $E_p$ could either indicate a gradually decreasing efficiency of the shocks in particle acceleration (i), or be due to a time decreasing Doppler boosting (ii). Figure~\ref{fig:toy_model} illustrates both possibilities.
The dip observed in the light curve is not neatly reproduced by our toy models. {\citet{Yi25b} interpreted the dip as a gap between two fast mini-jet pulses, implying that GRB\,230307A's time profile consists solely of short pulses without a slow variability component. Our model builds on the same assumption and also produces dips, most of which are less pronounced than the observed one, though. Alternative explanations include a temporary shutdown of the central engine or absorption/geometrical blocking along the line of sight. Given the rarity of this feature (possibly unique), stochasticity cannot be ruled out as its cause}.

An alternative scenario invokes a central engine emitting a single shell, that is then expanding. In this scenario, the GRB emission could be due to: (a) an external shock between the expanding shell and inhomogeneities (clumps) present in the circumburst medium \citep{Fenimore96,Dermer99,Dermer08}, or (b) magnetic reconnection events happening within the expanding shell, as proposed by \citet{Yi25}.
In both cases, the late, soft, and broad emission could be due to high-latitude photons arriving from large angles with respect to the line of sight. A major drawback of (a) is that the rate of shocked clumps, under the simple assumption of a spatially homogeneous distribution of clumps, should increase with time, whereas a decreasing rate of pulses is observed. The only way out would be assuming that the clumps are spatially clustered around the jet axis, which should roughly coincide with the line of sight,— a rather contrived assumption.

In (b) the emission arises from magnetic reconnection within a single expanding shell. A brief energy injection from the central engine triggers turbulence near the jet axis, which then spreads across the shell, producing delayed emission in expanding rings. This process leads to a broad pulse and naturally explains the observed spectral softening and the softer-wider/softer-later behaviour seen in GRBs like GRB\,230307A.

\begin{table*}[h!]
\centering
\scalebox{0.75}{
\begin{tabular}{ccccc}
\hline
\hline
\\
& Observed properties & Single shell model & Multiple shells model\\
\\
&Evolution of the peak rates & FRED shape expected & FRED envelope formed as a superposition of the shots \\
& Evolution of the peak times/waiting times & constrained spatial distribution of the clumps & fading central engine activity \\
&Evolution of the FWHM &  high latitude emission and increasing emission radius & slower late emitted shells \\
&Evolution of the peak energy & high latitude emission and increasing emission radius & slower late emitted shells \\
\\
\hline
\hline
\end{tabular}
}
\label{tab:summary_interpretations}
\caption{Summary of the different theoretical interpretations of the observed temporal and spectral properties of GRB\,230307A.}
\end{table*}

\section{Summary and conclusions}
\label{sec:conc}
We discovered new features in the GRB\,230307A time profile using exquisite data from GECAM. In particular, the waiting times, peak rate, pulse FWHM, and peak energy follow a characteristic exponential evolution over time. We show that these properties are also present in other long GRBs, that are merger candidates, suggesting the existence of a new subclass of long GRBs—so-called type IL—as proposed in the literature \citep{Wang25,Tan25}. These would originate from compact object mergers rather than from collapsars, as is typical for classical long GRBs. 
We built a toy model that is able to accurately reproduce GRB\,230307A light curve, which we optimised using a genetic algorithm. We proposed different theoretical scenarios to explain the observed trends, assuming either a central engine emitting multiple shells that collide with previously ejected material, or a single shell that expands and dissipates its energy progressively at larger radii. We also revisit the idea that external/refreshed shocks could explain prompt emission variability. While this scenario was previously disfavoured due to the apparent lack of timescale evolution within a sample of BATSE bursts, we show that the main criticisms that were historically used to support internal shocks as opposed to an external shocks interpretation of the GRB prompt emission, do not apply to GRB\,230307A, whose FWHMs clearly increase over time.

{In summary, while our toy and physical models are attempts to account for the unprecedented observed phenomenology within self-consistent pictures, alternative interpretations cannot be excluded in principle. Yet the fact remains that GRB\,230307A, along with other long compact object mergers candidates, exhibits a set of properties that is rarely seen in any other long GRBs, especially in GRBs with associated SNe that are currently known. Such a potentially distinctive signature of this class of GRBs may help their identification and challenges our current understanding of how GRB central engines operate.}

Identifying clear prompt $\gamma$-ray signatures of events coming from compact object mergers is particularly relevant in an era where new space missions such as the \textit{Space-based multi-band astronomical Variable Objects Monitor} (SVOM; \citealt{SVOM22}) and \textit{Einstein Probe} \citep{EinsteinProbeMission} are capable of performing prompt X-ray and optical follow-up of merger candidates, consequently enhancing the chances of coincident gravitational wave detection.


\section*{Acknowledgements}
{We are grateful to the anonymous reviewer for their valuable report, which helped us improve the quality of this work.}
R.M. and M.M. acknowledge the University of Ferrara for
the financial support of their PhD scholarships. M. B. acknowledges the Department of Physics and Earth Science of the University of Ferrara for the financial support through the FIRD 2024 grant.

\appendix

\section{Optimisation of the model}
\label{sec:model_optim}

To optimise our toy model, we employed a genetic algorithm (GA), whose principle is described in \citet{Bazzanini24}. We are using four losses to optimise the different models we considered: 
\begin{itemize}
    \item A loss $L_{\rm{avgd}}$ regarding the averaged time profile of GRB\,230307A. To compute $L_{\rm{avgd}}$, we applied a $5~{\rm s}$ window moving averaged to the simulated and observed time profile and have computed \begin{equation}
    \label{eq:loss_avgd}
    L_{avgd} =\sum_{i=1}^{i=N_{\rm bins}}(y_{sim,i}-y_{obs,i})^{2},
    \end{equation}
    where $y_{sim,i}$ and $y_{obs,i}$ are the averaged observed and simulated time, {and $N_{\rm bins}$ is the number of bins of the 5 s binned time averaged profile}.
    \item A loss $L_{\rm{peak}}$ regarding the peak times distribution. $L_{\rm{peak}}$ is computed by performing a two-population Kolmogorov-Smirnov (KS) test between the simulated and the observed peak time distributions. If $p$ is the $p$-value of the KS test, \begin{equation}
    \label{eq:loss_peak}
        L_{\rm{peak}} = \begin{cases}
        1-\log(p)       & \text{if } p \leq 0.05, \\
0        & \text{otherwise}.
        \end{cases}
    \end{equation}
    \item A loss $L_{N_p}$ regarding the number of peaks produced by the model. \begin{equation}
        L_{N_p} = 3\Big|\log{\Big(\frac{N_{\rm{obs}}}{N_{\rm{sim}}}\Big)}\Big|.
    \end{equation} The number of peaks in the observed and simulated profiles were obtained by applying {\sc mepsa} to these light curves. We defined $L_{N_p}$ so to have $L_{N_p}=0$ when $N_{\rm{obs}}=N_{\rm{sim}}$ and growing when the ratio deviates from 1, the absolute value is chosen to equally penalise the solution with too few or too many simulated peaks. The factor 3 is here to ensure that $L_{N_p}$ weights equally as the other losses.
    \item A loss $L_{\rm{FWHM}}$ regarding the FWHM distribution. The FWHMs of the pulses contained in the simulated and observed profiles were computed using {\sc mepsa}, and the method described in \citet{Camisasca23}. $L_{\rm{FWHM}}$
is computed by performing a two-population KS test between the simulated and the observed FWHM distributions, and computed similarly to $L_{\rm{peak}}$. 
    \end{itemize}
We finally took the average of the four losses described above to compute the total loss $L_{\rm{tot}}$. Each individual’s set of parameters was initially sampled randomly from a uniform distribution in the logarithmic space of each quantity, within the following ranges:

\begin{itemize}
    \item $N_0         	\in 	 [100,2000]$.
    \item $\tau       	\in	[5,25]~s$.
    \item ${\rm FWHM_0}      \in 	[0.01,0.5]~s$.
    \item $\tau_{F}    \in 	[10,30]~s$.
    \item $\tau_{\rm{rise}}     	\in 	[0.1-100]  s.$
    \item $N_F     	\in 	[500,2000]$.
\end{itemize}

In addition, for the two physical models, $\Gamma$, $\Gamma_s$, and $\Gamma_0$ were sampled in $1-10^4$, $R_0$ in $10^{10}-10^{17}~{\rm cm}$ and $\tau_{\Gamma}$ in $1-10^3~{\rm s}$. These ranges were chosen to constrain the parameter space in a way that ensures coverage and diversity among potential solutions.
For each individual, the loss was computed using 100 light curves generated from the same parameter set. We ran the genetic algorithm on at least 60 generations, using a population of 2000 individuals. From the total population, 300 individuals were selected for mating in each generation. The probability for a random mutation of the parameters was set to  10\%. 

\subsection{Results of the toy models}
The final optimised parameters are given by the median values of the individual parameter distributions in the final generation. The uncertainties are defined by the 16th and 84th quantiles of these distributions. We generated 1000 light curves to test the best-fit parameters, reported in Table~\ref{tab:model_comparison}, for the three models considered in this study and computed the loss on these profiles. The values of the individual losses and the total losses for the toy model (Sect.~\ref{sec:toy_model}) and the physical model 1 (Sect.~\ref{sec:phys_model1}) and 2 (Sect.~\ref{sec:phys_model2}) are reported in Table~\ref{tab:loss_summary}.

\begin{table*}[ht]
\centering
\small
\setlength{\tabcolsep}{5pt}
\begin{tabular}{@{}lccc@{}}
\toprule
{Loss Component} & {Toy Model} & {Model 1} (Constant $\Gamma$) & {Model 2} (Declining $\Gamma$) \\
\midrule
$L_{\mathrm{Np}}$ (number of peaks) & 0.091 & 0.370 & 0.181 \\
$L_{\mathrm{peak}}$ (peak times)     & 0.000 & 0.000 & 0.000 \\
$L_{\mathrm{FWHM}}$ (FWHM dist.)     & 1.160 & 0.047 & 0.012 \\
$L_{\mathrm{avgd}}$ (smoothed profile) & 0.860 & 0.935 & 0.770 \\
\midrule
{Total loss $L_{\mathrm{tot}}$} & {0.527} & {0.340} & {0.241} \\
\bottomrule
\end{tabular}
\caption{Loss function values from the genetic algorithm optimization for the three models.}
\label{tab:loss_summary}
\end{table*}

\begin{figure}[h!]
    \centering
    \includegraphics[width=0.5\textwidth]{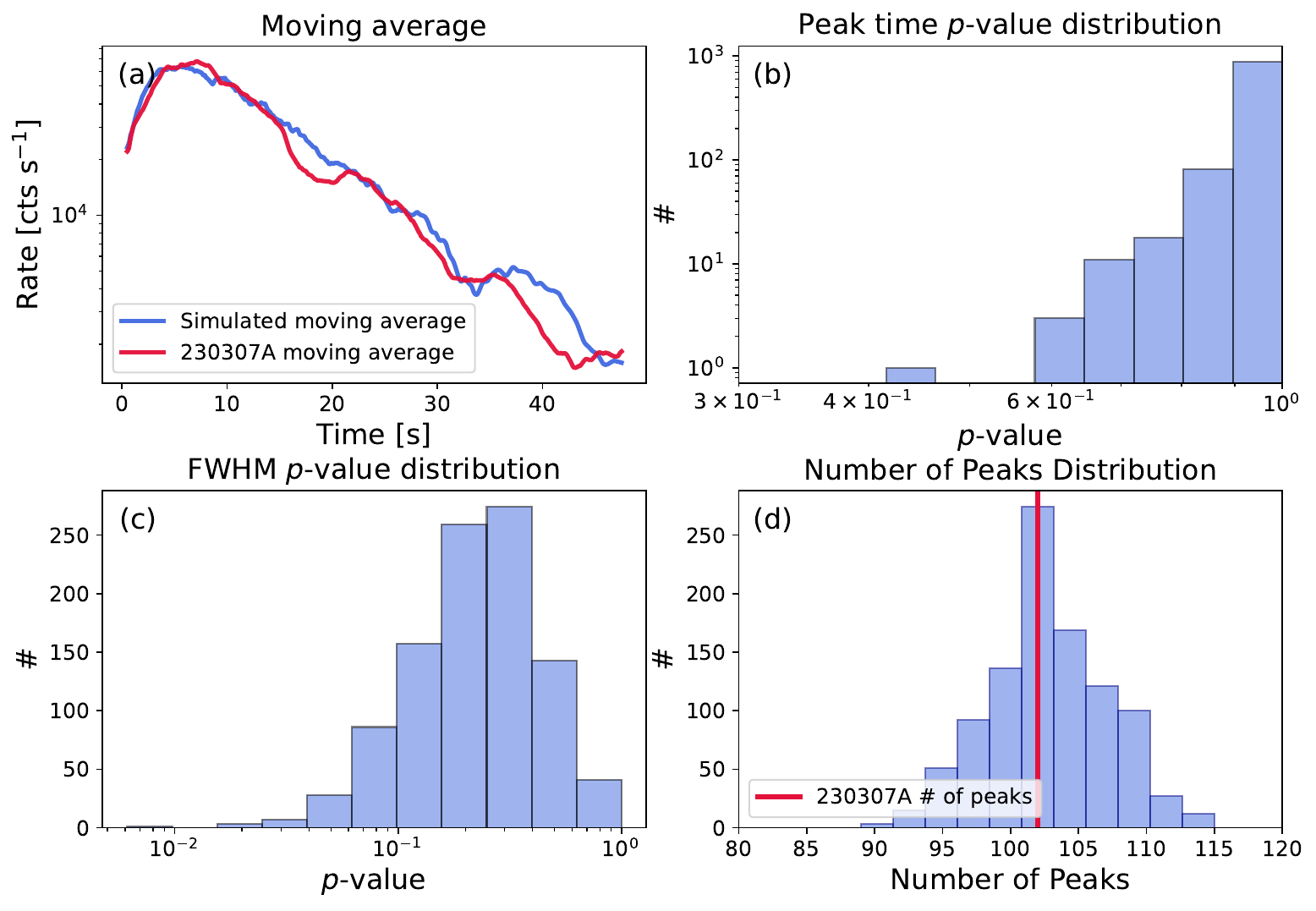}
    \caption{Four losses involved in the GA optimisation. Panel (a): Moving average time profile (5 s window) for the real (red) and simulated (blue) LCs.
Panel (b): Distribution of p-values from the KS test comparing peak time distributions.
Panel (c): Same as (b), but for FWHM distributions.
Panel (d): Number of peaks detected in simulated LCs (blue histogram) compared with GRB\,230307A (vertical red line).}
    \label{fig:4losses_plot}
\end{figure}

\begin{figure*}[h!]
    \centering
    \includegraphics[width=0.45\textwidth]{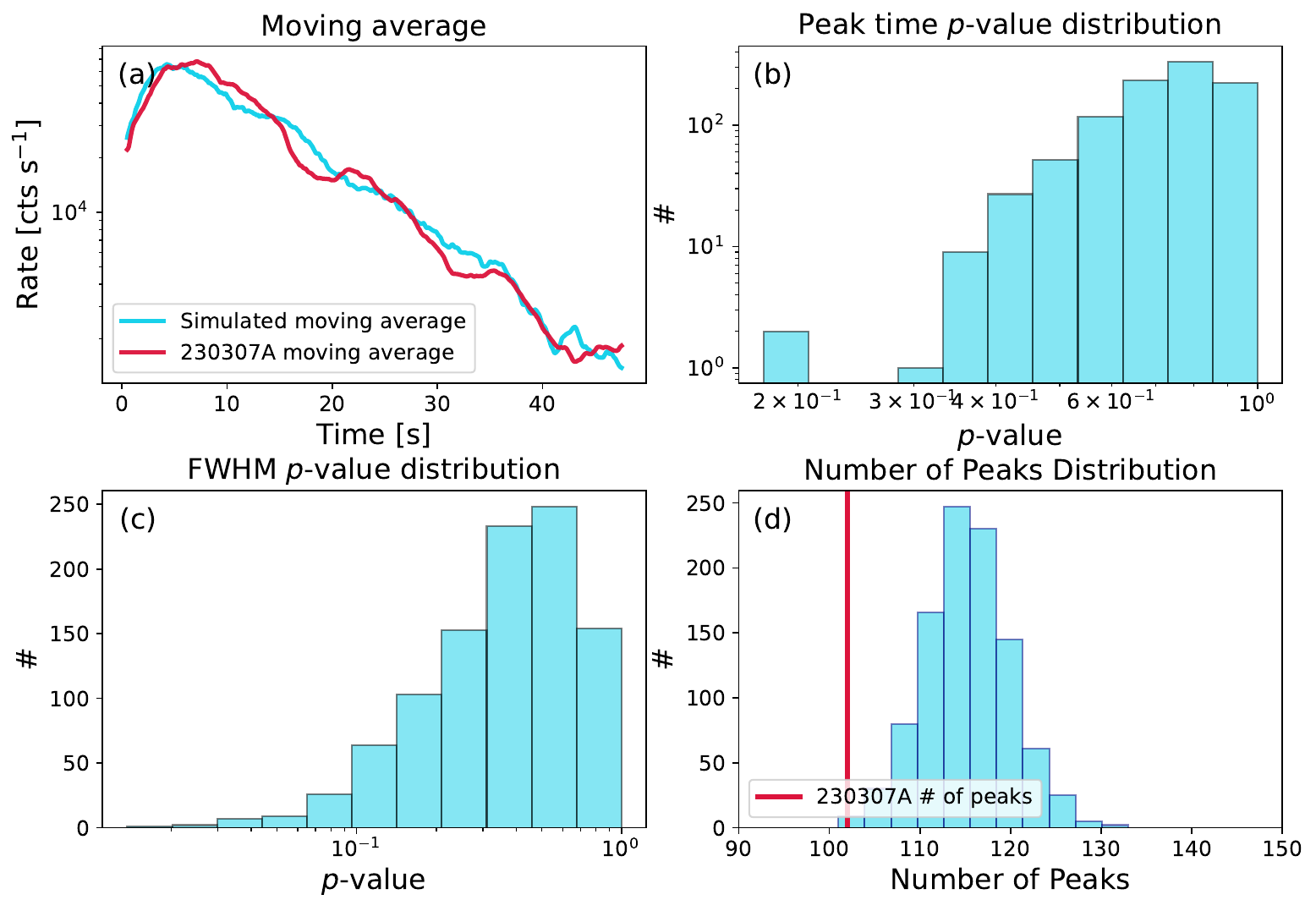}
    \includegraphics[width=0.45\textwidth]{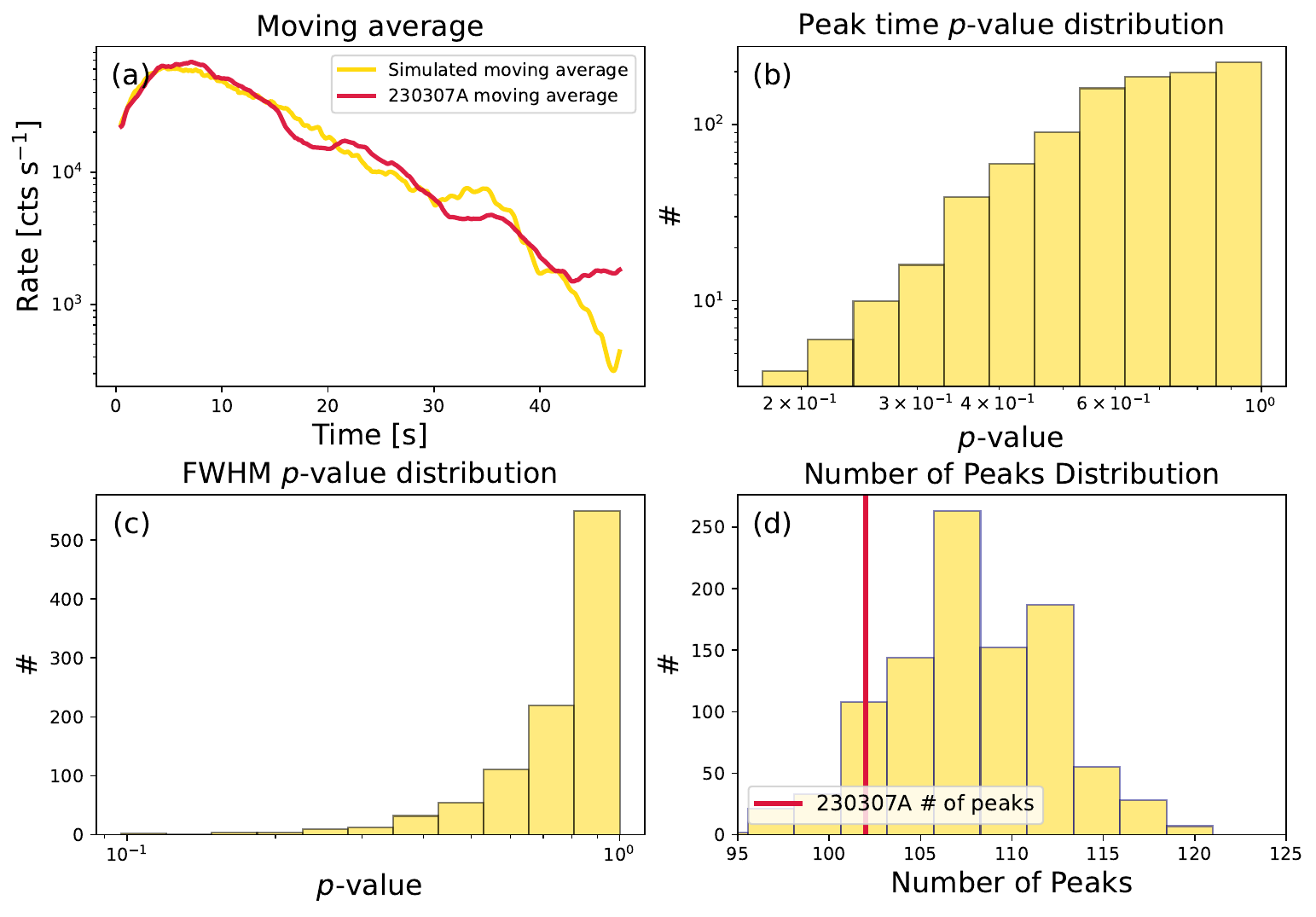}
    \caption{Same as in Fig.~\ref{fig:4losses_plot}, but for the constant (left panel) and decreasing (right panel) $\Gamma$-shell emission model.}
    \label{fig:4losses_plot2}
\end{figure*}

Figure~\ref{fig:4losses_plot} and \ref{fig:4losses_plot2} illustrate the outcome of the GA optimisation, showing how the synthetic light curves reproduce the main temporal features of GRB\,230307A, including the overall smoothed time profile, the distributions of peak times and FWHMs, and the total number of detected peaks.

\section{Analysis of other long merger candidates}
\label{sec:other_mergers}

{As anticipated in Section~\ref{sec:otherCOMs}, adopting the procedure of Section~\ref{sec:res}, we analysed the temporal behaviour of other long merger candidates, namely GRB\,211211A and GRB\,060614. Results are reported in Table \ref{tab:GRB_comp}.}

{Overall, we found that the exponential model provides an accurate description of the data (see $\chi^{2}$ values reported in Table~\ref{tab:GRB_comp}). In this respect, GRB\,211211A behaves very similarly to GRB\,230307A, consistent with previous claims (e.g. Peng et al. 2024). The WTs, the FWHMs, and the peak rates evolve with time as exponentials with timescales close to $20$ s. In the case of GRB\,060614, the different timescales are longer than in the case of GRBs\,211211A/230307A, and are less tightly constrained. In this case, both WT and FWHM increase, and the PR decay does not seem to share a common timescale; although, given the parameters' uncertainties, we cannot exclude the possibility that they are equal.}

\begin{table*}[ht]
\centering
\begin{tabular}{lccc}
\toprule
{Property} & {GRB\,211211A} & {GRB\,230307A} & {GRB\,060614} \\
\midrule
\multirow{4}{*}{Waiting times} 
    & $\tau = 22.7^{+6.3}_{-4.0}$ s & $\tau = 15.8^{+2.2}_{-1.8}$ s & $\tau = 125^{+291}_{-56}$ s \\
    & $N_0 = 136^{+23}_{-19}$ & $N_0 = 124^{+12}_{-11}$ & $N_0 = 68^{+106}_{-21}$  \\
    & $\sigma = 0.85^{+0.12}_{-0.10}$ & $\sigma = 0.54^{+0.07}_{-0.06}$ & $\sigma = 0.66^{+0.17}_{-0.12}$ \\
    & $\chi^2 = 86.5$ (dof = 86) & $\chi^2 = 100.1$ (dof = 99) & $\chi^2 = 32.4$ (dof = 32) \\
\midrule
\multirow{4}{*}{FWHM} 
    & $\tau_F = 21.6^{+4.4}_{-3.2}$ s & $\tau_F = 18.6^{+3.2}_{+2.5}$ s & $\tau_F = 74^{+97}_{-28}$ s \\
    & ${\rm FWHM_0} = 0.09^{+0.02}_{-0.02}$ s & ${\rm FWHM_0} = 0.12 \pm 0.02$ s & ${\rm FWHM_0} = 0.85^{+0.50}_{-0.32}$ s \\
    & $\sigma = 0.70^{+0.11}_{-0.10}$ & $\sigma = 0.48^{+0.09}_{-0.07}$ & $\sigma = 0.90^{+0.24}_{-0.18}$ \\
    & $\chi^2 = 86.6$ (dof = 87) & $\chi^2 = 99.9$ (dof = 100) & $\chi^2 = 32.4$ (dof = 33) \\
\midrule
\multirow{4}{*}{Peak rate} 
    & $\tau_p = 18.8^{+2.0}_{-1.6}$ s & $\tau_p = 10.9 \pm 0.6$ s & $\tau_p = 51.4^{+10.4}_{-7.4}$ s \\
    & $P_0 = 6.5^{+1.2}_{-1.0} \times 10^{4}$ cts/s  & $P_0 = 1.56^{+0.21}_{-0.18} \times 10^5$ cts/s & $P_0 = 1.33^{+0.30}_{-0.25}$ cts/s \\
    & $\sigma = 0.32^{+0.06}_{-0.05}$ & $\sigma = 0.22^{+0.05}_{-0.03}$ & $\sigma = 0.39^{+0.10}_{-0.07}$ \\
    & $\chi^2 = 56.1$ (dof = 56) & $\chi^2 = 48.7$ (dof = 49) & $\chi^2 = 33.5$ (dof = 33) \\
\bottomrule
\end{tabular}
\caption{{Results and comparison of exponential temporal evolution modelling of WTs, FWHMs, and peak rates for GRB\,211211A, GRB\,230307A, and GRB\,060614.}}
\label{tab:GRB_comp}
\end{table*}

\section{Analysis of SN-GRBs}
\label{sec: collapsar}
\begin{figure*}[ht]
    \centering
    \includegraphics[width=\linewidth]{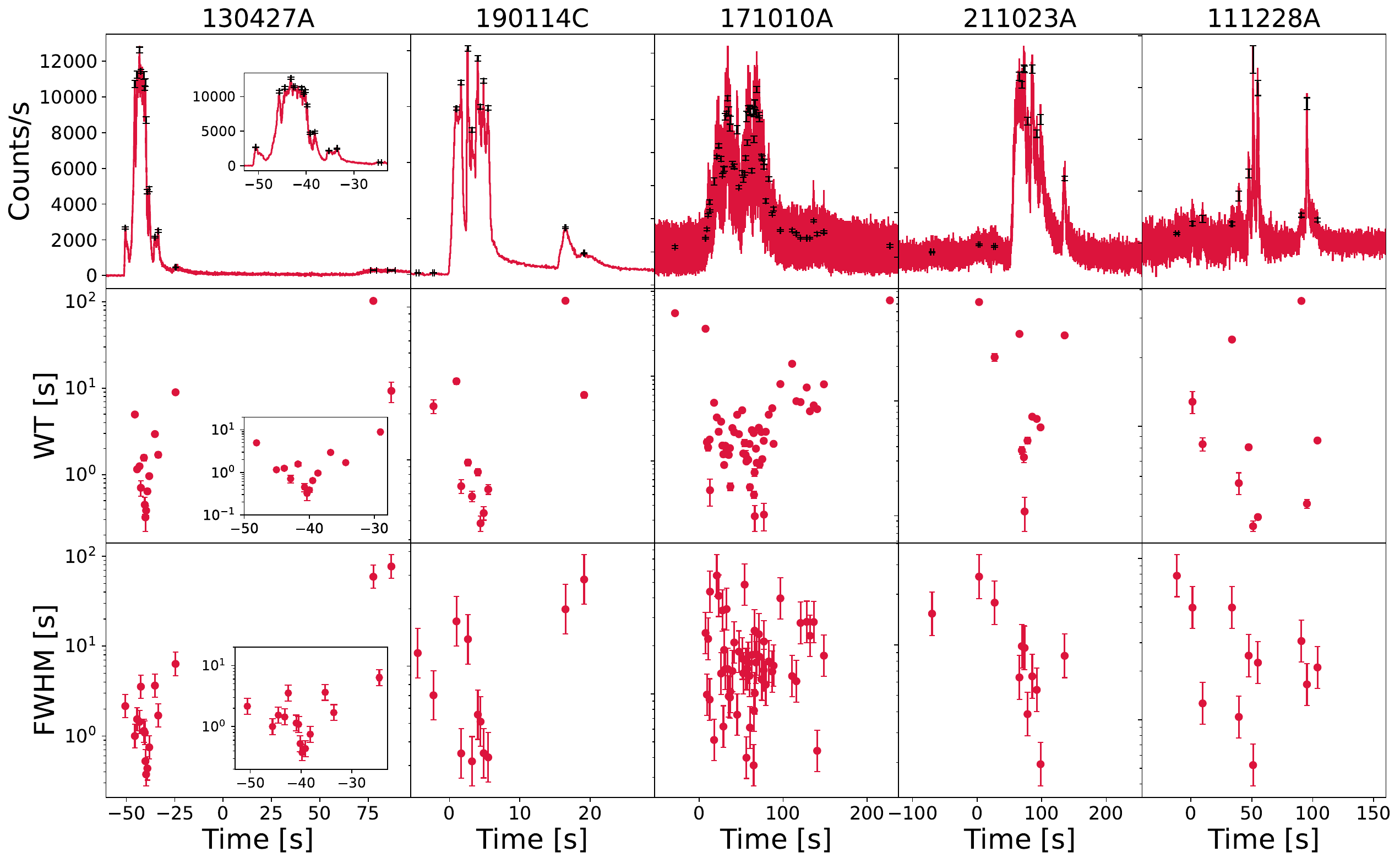}
    \caption{{ Five SN-GRBs having $N_p \geq 10$. Each top panel shows the $\gamma$-ray time profile (obtained with either \textit{Fermi}/GBM or \textit{Swift}/BAT). Black points represent the peaks detected by {\sc mepsa}. Middle and bottom panels respectively show the temporal evolution of WT and of FWHM throughout each burst. For GRB\,130427A, two insets show a close-in view of the densely populated interval.}}
    \label{fig:5SN_GRBs}
\end{figure*}

{
We investigated whether some Type-II GRBs could display behaviour similar to GRB\,230307A by focusing on bursts associated with Type Ic-BL supernovae, a clear indicator of a collapsar origin. Table~\ref{tab:SNGRB_peaks} reports a list of known SN-GRB associations, along with the corresponding number of detected peaks. We searched for SN-GRBs with sufficiently complex light curves ($N_p \geq 10$), as detected by {\sc mepsa} using either \textit{Swift}/BAT or \textit{Fermi}/GBM data (same dataset as \citealt{Guidorzi24}).  Only six GRBs met this criterion, namely GRB\,080319B, GRB\,111228A, GRB\,130427A, GRB\,171010A, GRB\,211023A, and GRB\,190114C (in addition to the exceptional GRB\,221009A, which is analysed separately). The LCs of these bursts, the corresponding WT and pulse FWHM evolution are displayed in  Fig.~\ref{fig:5SN_GRBs}. None of these GRBs exhibits a similar joint temporal evolution of the observables to that of GRB\,230307A.}

\begin{figure*}[ht]
    \centering
    \includegraphics[width=\linewidth]{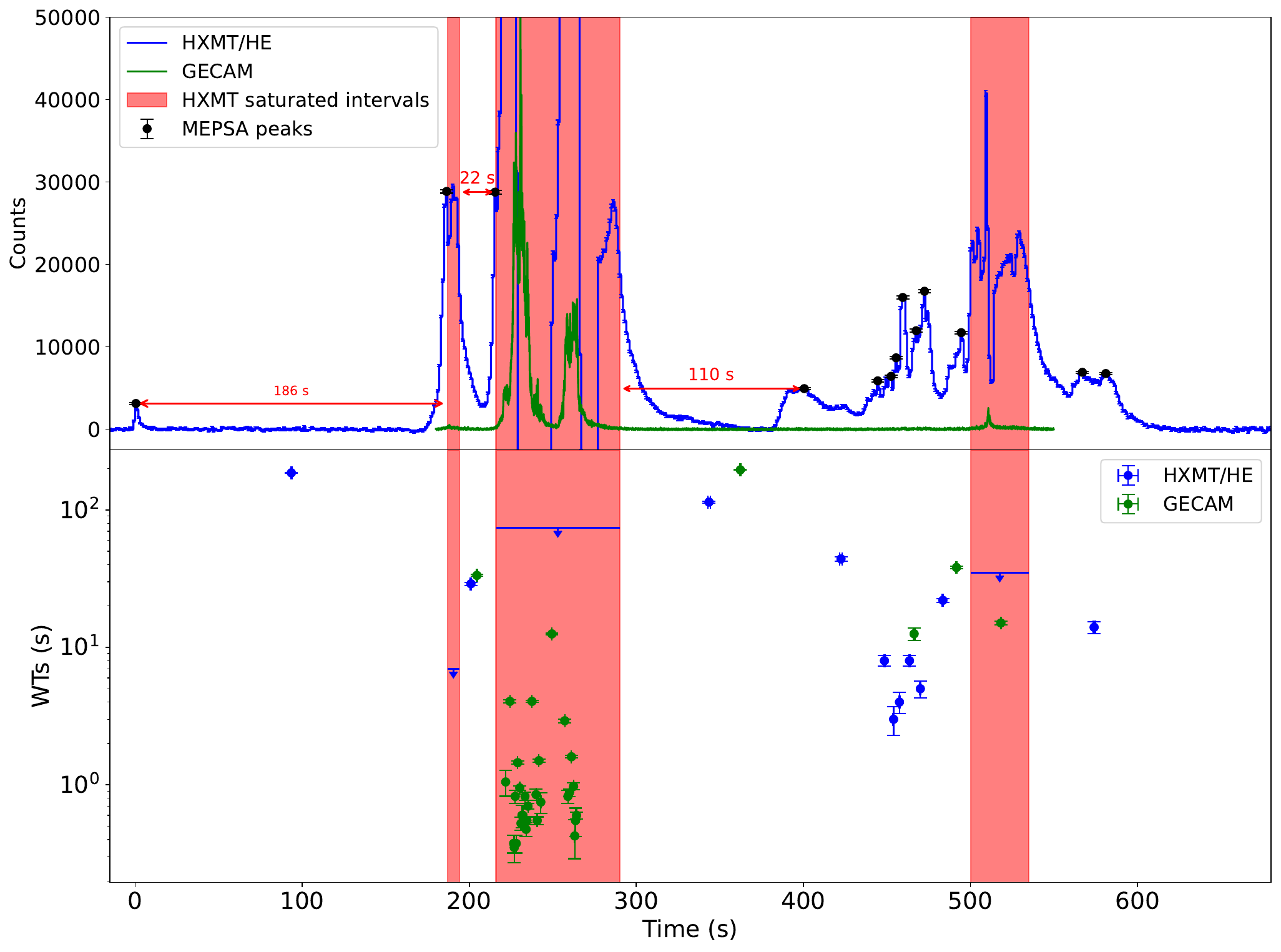}
    \caption{{ \textit{Top panel}: Background-subtracted LCs of GRB\,221009A from HXMT/HE (blue, 1 s binning) and GECAM (green, 50 ms binning). Red shaded areas mark intervals were HXMT/HE is affected by saturation. Peaks detected by {\sc mepsa} are shown as black markers. The two long quiescent intervals between the precursor and main emission (186 s) and between the main and late emission (110 s) are indicated. \textit{Bottom panel:} WTs derived from {\sc mepsa} detections in HXMT/HE and GECAM data as a function of time. WTs inside HXMT/HE saturated intervals are computed from GECAM measurements.}}
    \label{fig:221009A_hxmt_gecam}
\end{figure*}

{The case of GRB\,221009A was also considered. Its exceptional brightness makes it hard to obtain a complete time profile, unaffected by strong dead time and electronics saturation effects. We carried out a detailed analysis using both HXMT/HE and GECAM data \citep{ZhangWL25}. HXMT/HE data cover $T_{0}-141$ to $T_{0}+1800$ s, with three saturation intervals (187–194 s, 216–290 s, 500–535 s). Data are binned at 1 s, and background estimated via 8th-order polynomial interpolation. GECAM data (180–550 s) cover the main and the last emission episodes, are unaffected by saturation and binned at 50 ms. {\sc mepsa} was applied to both data sets, excluding the HXMT/HE saturated intervals. The combined HXMT/HE and GECAM light curves, along with the derived WTs, are shown in Fig.~\ref{fig:221009A_hxmt_gecam}. The complex profile consists of two long quiescent times: one between the precursor and the main emission, and the other between the main and late emission, separated by closely spaced pulses during each outburst. WTs in the main and late emission episodes were derived from GECAM data. Apart from  noting that the WTs belonging to the late emission are somehow longer than those of the main episode, the overall evolution of WTs with time is very different from the monotonic WT rise seen in GRB\,230307A.}

\begin{table}[ht]
\centering
\begin{tabular}{l l c c c}
\hline
GRB & SN & SN Spec & $z$ & $N_p$ \\
\hline
221009A & 2022xiw & 1 & 0.1505 & - \\
171010A & 2017htp & 1 & 0.3285 & 49 \\
080319B & ------ & 0 & 0.937 & 43 \\
130427A & 2013cq & 1 & 0.3399 & 16 \\
211023A & ------- & 1 & 0.39 & 12 \\
111228A & ------ & 0 & 0.71627 & 11 \\
190114C & 2019jrj & 1 & 0.4245 & 11 \\
090618 & ------ & 0 & 0.54 & 9 \\
111228A & ------ & 0 & 0.71627 & 6 \\
140506A & ------ & 0 & 0.889 & 5 \\
091127 & 2009nz & 1 & 0.49034 & 4 \\
140506A & ------ & 0 & 0.889 & 3 \\
060729 & ------ & 0 & 0.54 & 3 \\
091127 & 2009nz & 1 & 0.49034 & 3 \\
180728A & 2018fip & 1 & 0.117 & 2 \\
190829A & 2019oyw & 1 & 0.0785 & 2 \\
140606B & iPTF14bfu & 1 & 0.384 & 2 \\
130831A & 2013fu & 1 & 0.4791 & 2 \\
100316D & 2010bh & 1 & 0.0592 & 1 \\
101219B & 2010ma & 1 & 0.5519 & 1 \\
130215A & 2013ez & 1 & 0.597 & 1 \\
130702A & 2013dx & 1 & 0.145 & 1 \\
101219B & 2010ma & 1 & 0.5519 & 1 \\
120422A & 2012bz & 1 & 0.283 & 1 \\
120714B & 2012eb & 1 & 0.3984 & 1 \\
120729A & ------ & 0 & 0.8 & 1 \\
171205A & 2017iuk & 1 & 0.0368 & 1 \\
141004A & ------ & 0 & 0.573 & 1 \\
161219B & 2016jca & 1 & 0.1475 & 1\\
200826A & ------- & 0 & 0.7481 & 1 \\
201015A & ------- & 1 & 0.426 & 1 \\
980425 & 1998bw & 1 & 0.0085 & 1 \\
\hline
\end{tabular}
\caption{SN-GRBs with number of peaks $N_p$ detected by {\sc mepsa}, using LCs binned at 64 ms. The SN Spec column indicates whether the evidence for the SN was both photometric and spectroscopic (1) or just photometric (0).}
\label{tab:SNGRB_peaks}
\end{table}

%
\bibliographystyle{elsarticle-harv} 
\bibliography{alles_grbs}






\end{document}